\documentclass{jfm}

\usepackage{cite}
\usepackage{url}

\usepackage{graphicx}
\usepackage{natbib}
\usepackage{amssymb,amsmath}

\newcommand{\C}{\mathbb C}
\newcommand{\e}{\eqref}

\ifCUPmtlplainloaded \else
  \checkfont{eurm10}
  \iffontfound
    \IfFileExists{upmath.sty}
      {\typeout{Found AMS Euler Roman fonts on the system,
                   using the 'upmath' package.}%
       \usepackage{upmath}}
      {\typeout{Found AMS Euler Roman fonts on the system, but you
                   dont seem to have the}%
       \typeout{'upmath' package installed. JFM.cls can take advantage
                 of these fonts, if you use 'upmath' package.}%
       \providecommand\upi{\upi}%
      }
  \else
    \providecommand\upi{\upi}%
  \fi
\fi

\ifCUPmtlplainloaded \else
  \checkfont{msam10}
  \iffontfound
    \IfFileExists{amssymb.sty}
      {\typeout{Found AMS Symbol fonts on the system, using the
                'amssymb' package.}%
       \usepackage{amssymb}%
       \let\le=\leqslant  
       \let\ge=\geqslant  
      }{}
  \fi
\fi

\ifCUPmtlplainloaded \else
  \IfFileExists{amsbsy.sty}
    {\typeout{Found the 'amsbsy' package on the system, using it.}%
     \usepackage{amsbsy}}
    {\providecommand\boldsymbol[1]{\mbox{\boldmath $##1$}}}
\fi

\newcommand{\pD}{\partial }

\newcommand\Real{\mbox{Re}} 
\newcommand\E{\mbox{e}}
\newcommand\I{\mbox{i}}
\newcommand\D{\mbox{d}}
\newsavebox{\astrutbox}
\sbox{\astrutbox}{\rule[-5pt]{0pt}{20pt}}

\newcommand\p{\ensuremath{\partial}}

\title[Branch cuts of Stokes wave on deep water. Part I
]{Branch cuts  of Stokes wave on deep water. Part I: Numerical solution and Pad\'e approximation}

\author[S.\,A.~Dyachenko,  P.\,M.~Lushnikov, and A.\,O.~Korotkevich]%
{Sergey A. Dyachenko$^{1,2}$,\ns
Pavel M. Lushnikov$^{3,4}$\thanks{Email address for correspondence: plushnik@math.unm.edu},\break and Alexander O.
Korotkevich$^{3,4}$}

\affiliation{$^1$Department of Mathematics, University of Illinois at Urbana-Champaign, 1409 W. Green Street,
Urbana, IL 61801, USA\\[\affilskip]
$^2$Department of Mathematics, University of Arizona, 617 N.~Santa Rita Ave., P.O. Box 210089,
Tucson, AZ 85721, USA\\[\affilskip]
$^3$Department of Mathematics and Statistics, University of New Mexico, Albuquerque, MSC01 1115, NM, 87131, USA\\[\affilskip]
$^4$Landau Institute for Theoretical Physics, 2 Kosygin Str., Moscow, 119334, Russia}

\begin{document}

\maketitle

\begin{abstract}
Complex analytical structure of Stokes wave for two-dimensional
potential flow of the ideal  incompressible fluid with  free surface
and infinite depth is analyzed. Stokes wave  is  the fully nonlinear
periodic gravity wave propagating with the constant velocity.
Simulations with the quadruple   (32 digits) and variable precisions
(more than 200 digits) are performed to find Stokes wave with high
accuracy and study the Stokes wave approaching its limiting form
with $2\upi/3$ radians angle on the
crest.   
A conformal map is used which maps a free fluid surface  of Stokes
wave into the real line with fluid domain mapped into the lower
complex half-plane. The Stokes wave is fully characterized by the
complex singularities in the upper complex half-plane.  These
singularities are addressed by rational (Pad\'e) interpolation of
Stokes wave in the complex plane.  Convergence of Pad\'e
approximation to the density of complex poles with the increase of
the numerical precision and subsequent increase of the number of
approximating poles reveals that the only singularities of Stokes
wave are branch points connected by branch cuts. The converging
densities are the jumps across the branch cuts. There is  one branch
cut per horizontal spatial period $\lambda$ of Stokes wave. Each
branch cut extends strictly vertically above the corresponding crest
of Stokes wave up to complex infinity. The lower end of branch  cut
is the square-root branch point located at the distance $v_c$ from
the real line corresponding to the fluid surface in conformal
variables.
The increase of the scaled wave height $H/\lambda$ from the linear
limit $H/\lambda=0$ to the critical value $H_{max}/\lambda$ marks
the transition from the limit of almost linear wave to a strongly
nonlinear limiting Stokes wave (also called by the Stokes wave of the greatest height). Here $H$ is the wave height  from the crest to the trough in physical variables.  The limiting Stokes wave emerges as the singularity
reaches the fluid surface. Tables of Pad\'e approximation for Stokes waves of different heights are provided.
These tables allow to recover the Stokes wave with the relative accuracy of at least $10^{-26}$.
The tables use from several poles for near-linear Stokes wave up to about  hundred poles to  highly nonlinear Stokes wave with $v_c/\lambda\sim 10^{-6}.$
\end{abstract}


\section{Introduction}\label{sec:introduction}


Theory of spatially periodic progressive (propagating with constant
velocity without change of the shape and amplitude) waves in two-dimensional
(2D) potential flow of an ideal incompressible fluid with free
surface in gravitational field was founded in pioneering works by~\citet{Stokes1847,Stokes1880} and developed further by~\citet{Michell1893},
\citet{Nekrasov1921,Nekrasov1951}, and many others (see e.g.
a book by~\citet{Sretenskii1976} for review of older works
as well as
\citet{MalcolmGrantJFM1973LimitingStokes,SchwartzJFM1974,Longuet-HigginsFoxJFM1977,Longuet-HigginsFoxJFM1978,Williams1981,WilliamsBook1985,TanveerProcRoySoc1991,CowleyBakerTanveerJFM1999,Longuet-HigginsWaveMotion2008,BakerXieJFluidMech2011.pdf}
and references there in for more recent progress). There are two major
approaches to analyze the Stokes wave, both originally  developed by
Stokes. The first approach is the perturbation expansion in
amplitude of Stokes wave called by the Stokes expansion. That
approach is very effective for small amplitudes but converges very
slowly (or does not converge at all, depending on the formulation according to~\citet{DHT1992})
as the wave approaches to the maximum height $H=H_{max}$ (also called
by the Stokes wave of the greatest height or the limiting Stokes wave). Here the height $H$ is defined at the vertical distance from the crest to the trough of
Stokes wave over a spatial period $\lambda$. The second approach is
to consider the limiting Stokes wave, which is the progressive wave
with the highest nonlinearity. Using conformal mappings Stokes found that the limiting Stokes
wave has the sharp angle of $2\upi/3$ radians on the
crest~\citep{Stokes1880sup}, i.e. the surface is non-smooth (has a
jump of slope) at that spatial point. That corner singularity
explains a slow convergence of Stokes expansion as $H\to H_{max}.$ The global existence of the limiting Stokes wave was proven by \citet{Toland1978OnExistenceOfLimitingWave} however lacking a proof  of a Stokes conjecture that the the
jump of the slope at the crest is exactly $2\upi/3$ radians. The Stokes conjecture was later independently proven by \citet{Plotnikov1982} and \citet{AmickFraenkelTolandActaMath1982}. 

\begin{figure}
\centering
\includegraphics[width=5in]{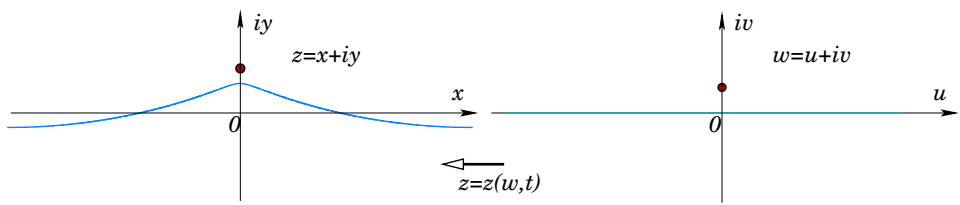}
\caption{\label{conformal_map} Schematic of a conformal map
between  the domain  below the solid  curve (left panel) in $z=x+\I y$
complex plane and the lower complex half-plane in $w=u+\I v$ (right panel).  Fluid
occupies the domain below the solid curve in physical plane
$z=x+\I y.$  The solid curve of  left panel (corresponds to a free
surface of the fluid) is mapped into the real line (another solid
line) in  right panel. One spatial period of Stokes wave is shown by
solid lines in both panels in the reference frame moving with the
velocity $c$.  The dark circles mark the positions of the
singularity closest to the fluid surface in both panels. }
\end{figure}

It was~\citet{Stokes1880sup} who first proposed to use conformal
mapping in order to address
 finite amplitude progressive waves.
In this paper we consider a particular case of potential flow of
the ideal fluid of infinite depth although  more general case of
fluid of arbitrary depth can be studied in a similar way. Assume
that  free surface is located at $y=\eta(x,t)$, where $x$ is the
horizontal coordinate, $y$ is the vertical coordinate, $t$ is the
time and $\eta(x,t)$ is the surface elevation with respect to the
zero mean level of fluid, i.e. $\int^\infty_{-\infty}\eta(x,t)\D x=0$. We
consider the conformal map between the domain
$-\infty<y\le\eta(x,t), \ -\infty<x<\infty$  of the complex plane
$z\equiv x+\I y$ filled by the infinite depth fluid and a lower
complex half-plane (from now on denoted by $\mathbb{C^-}$) of a
variable $w\equiv u+\I v$ (see Fig.~\ref{conformal_map}). The real
line $v=0$ is mapped into the free surface by $z(w)$ being the
analytic function in the lower half-plane of $w$ as well as the
complex fluid velocity potential $\Pi(w)$ is also analytic in
$\mathbb{C^-}$.
Both $z(w)$ and $\Pi(w)$ have singularities in upper half-plane
(here and further denoted by $\mathbb{C}^+$).

The  knowledge of
singularities in $\mathbb{C}^+$ would result in the efficient
description of the solution in the physical variables. Examples of
such type of solutions in hydrodynamic-type systems are numerous
including e.g. the dynamics of free surface of ideal fluid with
infinite depth~\citep{TanveerProcRoySoc1993,KuznetsovSpektorZakharovPhysLett1993,KuznetsovSpektorZakharovPRE1994} and finite
depth~\citep{DyachenkoZakharovKuznetsovPlasmPhysRep1996}, dynamics
of interface between two ideal
fluids~\citep{KuznetsovSpektorZakharovPhysLett1993}, ideal fluid
pushed through viscous fluid in a narrow gap between two parallel
plates (Hele-Shaw flow)~\citep{MineevZabrodinPRL2000},  the
dynamics of the interface between ideal  fluid and light viscous
fluid \citep{LushnikovPhysLettA2004} and bubble pinch-off \citep{TuritsynLaiZhangPRL2009}. In these systems the
dynamics is determined by  poles/branch cuts in the complex plane. Related systems
correspond to  the spontaneous appearance of curvature singularities on vortex
sheets as obtained by \citet{MooreProcRSocLond1979}.
Nie \& Baker (1998)  established that Moore's singularities are
present in axisymmetric vortex sheets. \cite{InogamovOparinJETP2003} considered cone-shaped nose of $2\upi/3$ degrees in
axisymmetric flow.  Ishihara \& Kaneda (1994) and Hou \&
Hu (2003) extended Moore's singularities to three-dimensional (3D) vortex sheets. Opposite limit is the global existence of water waves for small enough data shown both for 2D  \citep{WuInventMath2009}
and 3D
\citep{WuInventMath2011}
flows.

In this paper we determine that for Stokes wave the lowest
singularities in  $\mathbb{C}^+$ of both $z(w)$ and $\Pi(w)$  are
the square-root branch points located   periodically at
$w=n\lambda+\I  v_c-ct, \ n = 0, \pm 1, \pm 2,...$  (we choose
the crests of Stokes wave to be located at $w=n\lambda-ct$) and we
determine $v_c$ numerically as a function of $H/\lambda$. Here $c$
is the velocity of propagation of Stokes wave  which depends on $H$.
In the previous work~\citet{DLK2013} we found that as $H\to H_{max}$, the branch point approaches real
axis with the scaling law
\begin{align} \label{vcscalinglaw}
v_c \propto (H_{max}-H)^{\delta},
\end{align}
where $\delta=1.48\pm 0.03$. We also provided an accurate estimation of maximum amplitude of the Stokes wave
$H_{max}$. Adiabatically slow approach of Stokes wave to its
limiting form during wave dynamics is one of the possible routes to
wave breaking and whitecapping, which are responsible for
significant part of energy dissipation for gravity
waves~\citep{ZKPR2007, ZKP2009}. Formation of a close to limiting
Stokes wave is also considered to be a probable final stage of
evolution of a freak (or rogue) waves in the ocean resulting in
formation of approximate limiting Stokes wave for a limited period
of time with following wave breaking and disintegration of the wave
or whitecapping and attenuation of the freak wave into wave of
regular amplitude
\citep{ZakharovDyachenkoProkofievEuropJMechB2006,RaineyLonguet-HigginsOceanEng2006}.

The paper is organized as follows. In Section
\ref{sec:Basicequations} we introduce the basic equations of 2D hydrodynamics in conformal conformal variables and reduce these equations to the equation for Stokes wave.  In Section
\ref{sec:NumericalStokes}  numerical approaches to simulation of Stokes wave are given together with the results of simulations. Also numerical procedures to recover the location and  type of the branch point  are discussed.
 Section
\ref{sec:PadeApproximationStokesWave} introduces a new variable $\zeta$ defining a second conformal transformation which maps  one spatial period of Stokes wave into the entire real line.  Then the Pad\'e approximation of Stokes wave is found in complex  $\zeta$ plane. The efficient  Alpert-Greengard-Hagstrom (AGH) algorithm \citep{AGH2000,LauClassQuantumGrav2004} is used to obtain the Pad\'e approximation. That algorithm allows to avoid the appearance of artificial zeros and poles of Pad\'e  approximation and achieves a spectral accuracy.  The convergence of the Pad\'e approximation to the branch cut singularity is established which allows to recover the jump at branch cut.   Section \ref{sec:Stokeswaveasintegral} relates jump at the branch cut in  $\zeta$ variable to the sum of periodically located branch cuts  in $w$  complex plane.
It is shown how to use the series expansion of the jump along branch cuts near branch points to recover the square-root singularity at the branch point. It is demonstrated that there are no more singularities in the finite complex plane beyond one branch point $w=\I v_c$ per period.  In Section \ref{sec:Conclusion} the main results of the paper are
discussed.
Appendix \ref{DynDeriv} provides a derivation of basic hydrodynamic equations in conformal variables. Appendix \ref{sec:AGHalgorithm} gives a short description of AGH algorithm adapted for Stokes wave. Appendix \ref{sec:TablesStokesWaves} describes a notation used for the  tables of Pad\'e approximants for Stokes wave and gives samples of such tables. A full set of tables is provided in the electronic attachment.
These tables  reproduce the Stokes wave with the relative accuracy of at least $10^{-26}$.
\section{Basic equations}
\label{sec:Basicequations}

 In physical coordinates $(x,y)$ a
velocity ${\bf v}$   of $2D$ potential flow of inviscid
incompressible fluid is determined by a velocity potential $\Phi(x,y,t)$ as
${\bf v}= \nabla \Phi$. The incompressibility condition $\nabla
\cdot {\bf v} = 0$ results in the Laplace equation
\begin{align} \label{laplace}
\nabla^2 \Phi = 0
\end{align}
inside fluid $-\infty<y<\eta(x,t)$. To obtain the closed set of
equations we add the decaying boundary condition at large depth
$\Phi(x,y,t)|_{y\to-\infty} = 0$, the kinematic boundary condition
\begin{align} \label{kinematic1}
\dfrac{\partial \eta}{\partial t}    =\left( -\dfrac{\partial
\eta}{\partial x}\dfrac{\partial \Phi}{\partial x}+
\left.\dfrac{\partial \Phi}{\partial y}\right)\right|_{y =
\eta(x,t)}
\end{align}
and the dynamic boundary condition
\begin{align} \label{dynamic1}
\left.\left(\dfrac{\partial \Phi}{\partial t} +
 \dfrac{1}{2}\left(\nabla \Phi\right)^2\right)\right|_{y = \eta(x,t)} + g\eta = 0
\end{align}
at the free surface
\begin{equation} \label{etadef}
y=\eta(x,t).
\end{equation}
 We  define the boundary value of
the velocity potential as
\begin{equation} \label{psidef}
\left.\Phi(x,y,t)\right|_{y = \eta(x,t)}
\equiv \psi(x,t),
\end{equation}

Consider a time-dependent conformal transformation
\begin{equation}\label{ztconformal}
 z=z(w,t), \quad w=u+iv
\end{equation}
which maps a half-strip $-\frac{\lambda}{2} \le u <
\frac{\lambda}{2}, \ -\infty < v \le0 $ of  complex plane $w$ into a
region $-\frac{\lambda}{2} \le x < \frac{\lambda}{2},  \ -\infty < y
\le \eta(x,t) $ of complex physical plane  $z=x+\I y$ at each time $t$
such that the line $-\frac{\lambda}{2} \le u < \frac{\lambda}{2}, \
v=0$ is mapped into a line of free surface $x+\I\eta(x,t)$ with
$-\frac{\lambda}{2} \le x < \frac{\lambda}{2}$ and

\begin{equation} \label{xL2}
x\left (-\frac{\lambda}{2}\right )=-\frac{\lambda}{2},  \quad x\left(\frac{\lambda}{2}\right )=\frac{\lambda}{2}.
\end{equation}
Also $w=-\I \infty$ maps into $z=-\I \infty$. Here the flow is assumed to be periodic in the horizontal direction with the period $\lambda$ both in $w$ and $z$ variables.
Conditions \eqref{xL2} suggest to separate $z(w,t)$ into a periodic part $\tilde z(w,t)$ and a non-periodic part $w$ as follows
\begin{equation}\label{xtildedef}
\begin{split}
& z(w,t)=w+\tilde z(w,t), \;\mathrm{or}\;
x(w,t)=u+\tilde x(w,t), \quad \tilde y(w,t)=v+y(w,t),
\end{split}
\end{equation}
where
\begin{align} \label{xtildeperiodic}
\tilde z(w+\lambda)=\tilde z(w), \quad \tilde x\left (\pm\frac{\lambda}{2}\right )=0.
\end{align}
Equations \eqref{xtildedef} and \eqref{xtildeperiodic} extend conformal transformation  \eqref{ztconformal}
into $\mathbb{C}^-.$ Also $x(u,t)$ and $y(u,t)$ form a parametric representation (over the parameter $u$) of the free surface elevation \e{etadef}.

 The idea of using time-dependent conformal transformation for unsteady fluid flow was  exploited by several authors including~\citet{Ovsyannikov1973},
 \citet{MeisonOrzagIzraelyJCompPhys1981}, \citet{TanveerProcRoySoc1991,TanveerProcRoySoc1993}, and \citet{DKSZ1996,ZakharovDyachenkoVasilievEuropJMechB2002}.
We follow \citet{DKSZ1996} to recast the system \eqref{laplace}-\eqref{dynamic1} into the equivalent form for $x(u,t), \ y(u,t)$ and $\psi(u,t)$ at the real line $w=u$ of the complex plane $w$ using the conformal transformation  \eqref{ztconformal} (see Appendix~\ref{DynDeriv} for more details).  A kinematic boundary condition \eqref{kinematic1} is reduced to %
\begin{equation}\label{fullconformal1}
 y_tx_u  -x_t  y_u + \hat H \psi_u = 0
\end{equation}
and the dynamic boundary condition \eqref{dynamic1} is given by
\begin{equation}\label{fullconformal2}
\psi_t y_u - \psi_u y_t + gyy_u =- \hat H \left (\psi_t x_u -
\psi_ux_t + gyx_u \right ),
\end{equation}
where
\begin{equation} \label{Hilbertdef}
\hat H f(u)=\frac{1}{\upi} \text{p.v.}
\int^{+\infty}_{-\infty}\frac{f(u')}{u'-u}\D u'
\end{equation}
is the Hilbert
transform with $\text{p.v.}$ meaning a Cauchy principal value of integral.  Periodicity of $f(u)$ allows to reduce the integration in the Hilbert transform as follows %
\begin{equation} \label{Hilbertdefperiodic}
\hat H f(u)=\frac{1}{\upi} \sum \limits _{n=\infty}^\infty \text{p.v.}
\int^{\lambda/2}_{-\lambda/2}\frac{f(u')}{u'-u+n\lambda}\D
u'=\frac{1}{\lambda} \text{p.v.}
\int^{\lambda/2}_{-\lambda/2}\frac{f(u')}{\tan{\left (
\upi\frac{u'-u}{\lambda}\right )}}\D u'.
\end{equation}

The equivalence of equations  \eqref{fullconformal1} and \eqref{fullconformal2} to equations  \eqref{laplace}-\eqref{dynamic1} uses the analyticity of $z(w)$ and $\Pi(w)$ in $\mathbb{C}^-$, where
\begin{equation} \label{ComplexPotentialdef}
\Pi=\Phi+\I \Theta
\end{equation}
is the complex velocity potential. Here $\Theta$ is the stream function defined by $\Theta_x=-\Phi_y$ and $\Theta_y=\Phi_x$ to satisfy Cauchy-Riemann conditions for analyticity of $\Pi(z,t)$ in $z$ plane.
The conformal transformation   \eqref{ztconformal} ensures that
\begin{align} \label{ThetauPhiv}
\Theta_u=-\Phi_v, \quad \Theta_v=\Phi_u
\end{align}
in $w$ plane.
The periodicity of the flow implies the condition %
\begin{align} \label{Piperiodic}
\Pi(w+\lambda,t)=\Pi(w,t)
\end{align}
together with equation \eqref{xtildeperiodic}.
We also assumed in  equations
\eqref{fullconformal1} and \eqref{fullconformal2} that%
\begin{equation} \label{yxucondition}
\int\limits^{\lambda/2}_{-\lambda/2}\eta(x,t)\D
x=\int\limits^{\lambda/2}_{-\lambda/2} y(u,t)x_u(u,t) \D u=0,
\end{equation}
meaning that the elevation of  free surface of unperturbed fluid
is set to zero. The equation \e{yxucondition} is valid at all times
and reflects a conservation of the total mass of fluid.

Both
equations \eqref{fullconformal1} and \eqref{fullconformal2} are defined on the
real line $w=u$. The Hilbert operator $\hat H$ transforms
into the multiplication operator
\begin{equation} \label{Hfk}
 (\hat H f)_k=\I\,
\text{sign}{\,(k)}\,f_k,
\end{equation}
for the Fourier coefficients (harmonics) $f_k$,
\begin{align} \label{ffourier}
f_k=\frac{1}{\lambda}\int\limits_{-\lambda/2}^{\lambda/2}
f(u)\exp\left (-\I ku\frac{2\upi}{\lambda}\right )\D  u,
\end{align}
 of  the periodic function $f(u)=f(u+\lambda)$ represented through the Fourier series
\begin{equation} \label{fkseries}
f(u)=\sum\limits_{k=-\infty}^{\infty} f_k\exp\left (\I
ku\frac{2\upi}{\lambda}\right ).
\end{equation}
Here $\text{sign}(k)=-1,0,1$ for $k<0, \ k=0$ and $k>0$, respectively.

The Fourier series \eqref{fkseries} allows to rewrite $f(u)=f(w)|_{v=0}$ as follows
\begin{equation} \label{fpm0}
f(u)=f^+(u)+f^-(u)+f_0,
\end{equation}
where
\begin{equation} \label{fplus}
f^+(w)=\sum\limits_{k=1}^{\infty} f_k\exp\left (\I
kw\frac{2\upi}{\lambda}\right )
\end{equation}
is the analytical function  in $\mathbb{C}^+$,
\begin{equation} \label{fminus}
f^-(w)=\sum\limits_{k=-\infty}^{-1} f_k\exp\left (\I
kw\frac{2\upi}{\lambda}\right )
\end{equation}
is the analytical function  in $\mathbb{C}^-$ and $f_0=const$ is the
zero harmonic of Fourier series \eqref{fkseries}. In other
words, equation \eqref{fpm0} decompose $f(u)$ into the sum of
functions $f^+(u)$ and $f^-(u)$ which are analytically continued
from the real line $w=u$  into $\mathbb{C}^+$ and $\mathbb{C}^-$,
respectively. Equations \e{Hfk}, \e{fpm0}, \e{fplus} and \e{fminus}
imply that
\begin{equation} \label{Hfdef}
\hat Hf(u)=\I \left[ f^+(u)-f^-(u)\right].
\end{equation}

 If function $f(w)$ is analytic in $\C^-$ then    $\bar{f}(\bar {w})$
 is analytic in $\C^+$ as follows from equations \e{fpm0}-\e{fminus}, where bar mean complex conjugation, $\bar w=u-\I  v$. Then the function $\bar {f}(u),$ $u\in\mathbb{R}$ has analytic continuation into $\C^+$ because at the real line $w=\bar w$.  Using equations \e{psidef} and \e{ComplexPotentialdef} we obtain that
$\psi(u,t)=\frac{1}{2}[\Pi(u,t)+\bar\Pi(u,t)]$. It means that after solving equations  \eqref{fullconformal1} and \eqref{fullconformal2} one can recover the complex potential $\Pi$ from the analytical continuation of
\begin{equation} \label{Picontinued}
\Pi(u,t)=2\hat P\psi(u,t)
\end{equation}
into
$\mathbb{C}^-$. Here
\begin{equation} \label{Projectordef}
\hat P=\frac{1}{2}(1+\I \hat H)
\end{equation}
is the projector operator, $\hat Pf=f^-+\frac{f_0}{2}$, into a
function which has analytical continuation from the real line $w=u$
into $\C^-$, as follows from equation \e{Hfdef}. Note that without
loss of generality we assumed the vanishing zero Fourier harmonic,
$\Pi_0=0,$ for $\Pi(u,t).$

Also
\begin{equation} \label{H2def}
\hat H^2f=-f
\end{equation}
for the function $f(u)$ defined by \e{fpm0} provided the additional restriction that $f_0=0$ holds.
In other words, the Hilbert transformation is invertible on the class of functions represented by their Fourier series
provided zeroth Fourier harmonic $f_0$ vanishes.
If $f_0\ne 0$ then the identity \e{H2def} is replaced by
\begin{equation} \label{H2def0}
\hat H^2f=-(f-f_0).
\end{equation}

 The analyticity of $z(w)$ in $\mathbb{C}^-$
implies, together with $\tilde x=\frac{1}{2}(\tilde z+\bar {\tilde z}) $,  $\tilde y=\frac{1}{2\I }(\tilde z-\bar {\tilde z})$ and equations \e{Hfdef},\e{H2def0}, that
at the real line $w=u$ the following relations hold%
\begin{align} \label{xytransform}
y-\tilde y_0=\hat H\tilde x \quad  \text{and} \quad  \tilde    x-\tilde x_0=-\hat Hy.
\end{align}
Here $\tilde x_0$ and $\tilde y_0$ are zero Fourier harmonics of
$\tilde x(u,t)$ and $y(u,t)$, respectively. Note that the addition of zero
harmonics  $\tilde x_0$ and $\tilde y_0$ into equation
\e{xytransform} is the modification compare with Refs.
\citet{DKSZ1996,ZDV2002}.  These Refs. were focused on the
decaying boundary conditions $\eta(x,t)\to 0$ and $\psi(x,t)\to 0$
for $|x|\to \infty$ which imply, together with the condition
\e{yxucondition} in the limit $\lambda \to \infty,$ that $\tilde
x_0=\tilde y_0=0$. However, generally  $\tilde x_0$ and $\tilde y_0$
might be nonzero for the periodic solutions with a finite $\lambda$
considered in this paper.

Equations \e{xytransform} imply that it is enough to find either $y(u,t)$ or $x(u,t)$ then the second of them is recovered by
these explicit expressions. Taking derivative of equations  \e{xytransform} with respect to $u$ results in the similar relations
\begin{align} \label{xytransform_der}
y_u=\hat H\tilde x_u \quad  \text{and} \quad  \tilde    x_u=-\hat Hy_u,
\end{align}

\subsection{Progressive waves}
\label{sec:Progressivewaves}

Stokes wave corresponds to a solution of the system
\eqref{fullconformal1} and \eqref{fullconformal2} in the traveling
wave form
\begin{align}\label{travelling}
\psi (u,t)= \psi(u-ct), \ \tilde z (u,t)= \tilde z(u-ct),
\end{align}
where both $\psi$ and $\tilde z$ are the periodic functions of
$u-ct$.  Here $c$ is the phase velocity of Stokes wave. We transform into the moving frame of reference, $u-ct\to
u$, and assume that the crest of the Stokes wave is located at $u=0$
as in Fig.~\ref{conformal_map} and $\lambda$ is the spatial period in $u$ variable
for  both $\psi$ and $\tilde z$ in equation \eqref{travelling}. We look for the  Stokes wave which has one crest per period. Higher order progressive waves are also possible which have more than one different peak per period \cite{ChenSaffmanStudApplMath1980}. However here we consider only Stokes wave.  We recall that the  spatial period $\lambda$  is the same in both $u$ and $x$ variables as follows from equation \e{xtildeperiodic}. In addition, it implies that  the phase velocity is the same both in  $u$ and $x$    variables so that the Stokes wave has the moving surface $y=\eta(x-ct)$ and the velocity potential $\psi=\psi(x-ct)$  in physical spatial variables $(x,y)$ with the same value of $c$ as in equations  \eqref{travelling}.  The Stokes
solution requires  $y(u)$ to be the even function while $\tilde
x(u)$ needs to be the odd function which ensures that $y=\eta(x-ct)$ is the even function.

It follows from \eqref{fullconformal1} and \eqref{travelling} (corresponding to substitution
$\frac{\p}{\p t}\rightarrow -c\frac{\p}{\p u}$ for $y$ and $\psi$ and $\frac{\p x}{\p t}\rightarrow -c\frac{\p \tilde x}{\p u}$) that
$ \hat H\psi_u =cy_u$ and then excluding $\psi$ from
\eqref{fullconformal2} we obtain that

\begin{equation}
\label{stokes_wave} -c^2y_u + gyy_u + g\hat H[y(1+\tilde x_u)] = 0.
\end{equation}
We now apply $\hat H$ to \eqref{stokes_wave}, use
\eqref{xytransform} to obtain a closed expression for $y$, and
introduce the operator   $\hat k \equiv -\partial_u \hat H =
\sqrt{-\nabla^2}$ which results in the following expression
\begin{equation}\label{stokes_wave2}
\begin{split}
& \hat L_0 y\equiv \left(  {c^2}\hat k - 1 \right) y -  \left( \frac{\hat k y^2}{2} + y\hat k y \right) = 0, \\
\end{split}
\end{equation}
where we made all quantities dimensionless by the following scaling
transform $u\to u\lambda/2\upi,\ x\to x\lambda/2\upi, \ y\to
y\lambda/2\upi $ and $c$ is scaled by $c_0$ as follows $c\to c \, c_0$,
where $c_0 = \sqrt{g/k_0}$ is the phase speed of linear gravity wave
with the wavenumber $k_0 = 2\upi/\lambda$. In these scaled units the
period of $\psi$ and $\tilde z$  is $2\upi$. Our new operator $\hat
k$ in Fourier space acts as multiplication operator, qualitatively similar to
$\hat H$: $(\hat k f)_k = k f_k$.

\section{Numerical simulation of Stokes wave}
\label{sec:NumericalStokes}

We solve  \eqref{stokes_wave2} numerically to find $y(u)$ by two
different methods each of them beneficial for different range of the
parameter $H/\lambda$. For both methods  $y(u) $ was expanded in
cosine Fourier series and the operator $\hat k$ was evaluated
numerically using Fast Fourier Transform (FFT). A uniform grid with
$M$ points was used for the discretization of $-\upi\le u< \upi$. A
first method is inspired by a Petviashvili
method~\citep{Petviashvili1976} which was originally proposed to
find solitons in nonlinear Schr\"odinger (NLS) equation as well as
it was adapted for nonlocal NLS-type equations, see e.g.
\citet{LushnikovOL2001}. We used a version generalized Petviashvili
method (GPM) \citep{LY2007,PelinovskyStepanyantsSIAMNumerAnal2004}
adjusted to Stokes wave as described in \citet{DLK2013}. In practice
this method allowed  to find high precision solutions  up to
$H/\lambda \lesssim  0.1388$. The performance of that method for
larger values of $H/\lambda$ was limited by the decrease of  the
speed of numerical  convergence.

\subsection{Newton CG and Newton CR methods}
\label{sec:NewtonCG}
For larger $H/\lambda$ we used a second method which
is the Newton Conjugate Gradient  (Newton-CG)
~method proposed by~\citet{JiankeYang2009,YangBook2010}. The idea behind the
Newton-CG method is simple and aesthetic: first, linearize
\eqref{stokes_wave2} about the current approximation $y_n$,
assuming that the exact solution can be written as a sum of current
approximation and a correction $y=y_n + \delta y_n$: $\hat L_0 y = 0. $ Then
$\hat L_0 y_n + \hat L_1\delta y_n   \simeq 0, $ where $\hat L_1=-\hat M\delta y_n-
\left( \hat k (y_n\delta y_n) + y_n\hat k \delta y_n+\delta y_n\hat k y_n
\right)$ is the linearization of $\hat L_0$ around the current approximation
$y_n$ and $\hat M\equiv{-c^2}\hat k + 1$. Second, solve the resulting linear system $\hat L_1 \delta y_n = -\hat L_0 y_n$
for $\delta y_n$
with one of  standard numerical methods, in our case it was
either Conjugate Gradient (CG) method~\citep{HS1952} or Conjugate Residual (CR)
method~\citep{Luenberger1970} to obtain next approximation
$y_{n+1}=y_n+\delta y$. It should be noted that monotonic
convergence of CG or CR methods is proven only for positive
definite (semidefinite for CR) operators, while in our case $\hat
L_1$ is indefinite. Nevertheless, both methods were converging (although generally nonmonotonically) to
the solutions, and convergence was much faster than using GPM.

Newton-CG/CR methods can be written in either Fourier space, or in
physical space. We considered both cases, however Newton-CG/CR
methods in Fourier space require four fast Fourier transforms per
CG/CR step, while in physical space it
requires at least six. 
For both cases CG and CR we used  $\hat M$ as a preconditioner.

We found that the region of convergence of the  Newton-CG/CR methods
to nontrivial physical solution  \eqref{stokes_wave2} is relatively (with respect to GPM)
narrow and requires  an initial guess $y_0$ to be quite close to
the exact solution $y$. In practice we first run GPM and then
choose $y_0$ for Newton-CG/CR methods as the last available iterate
of GPM.

Because most of our interest was in getting dependence of characteristics of Stokes waves on the wave height and the only
parameter in equation~\eqref{stokes_wave2} is velocity of propagation $c$, we were calculating waves changing continuously
the parameter $c$ and using results of computations with previous values of $c$ as initial condition $y_0$ for the
Newton CG/CR iterations.
Due to this approach Newton-CG/CR methods converge to the nontrivial solution in all cases provided we additionally used the numerical procedure described below in  Section \ref{sec:velocityvssteepness}.

\subsection{Stokes wave velocity as a function of steepness}
\label{sec:velocityvssteepness} 
\begin{figure}
\includegraphics[width=2.65in]{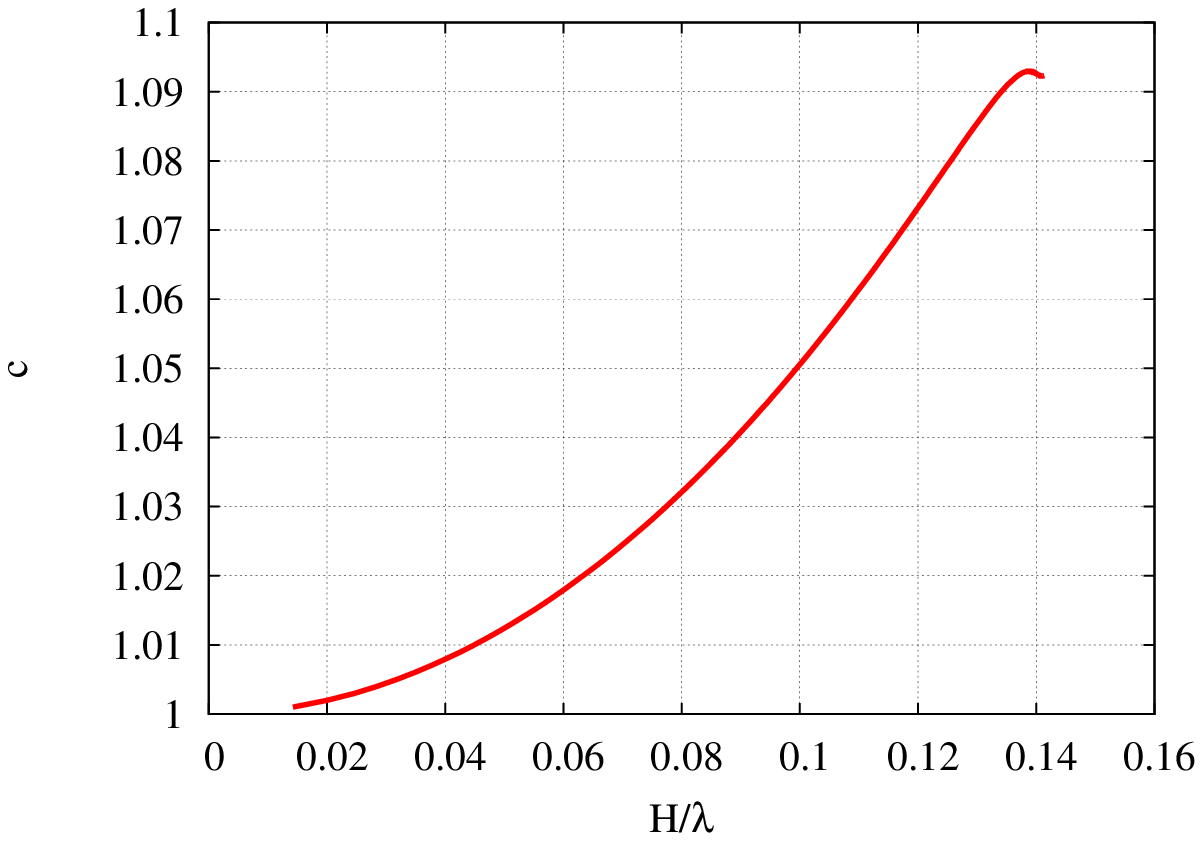}
\includegraphics[width=2.65in]{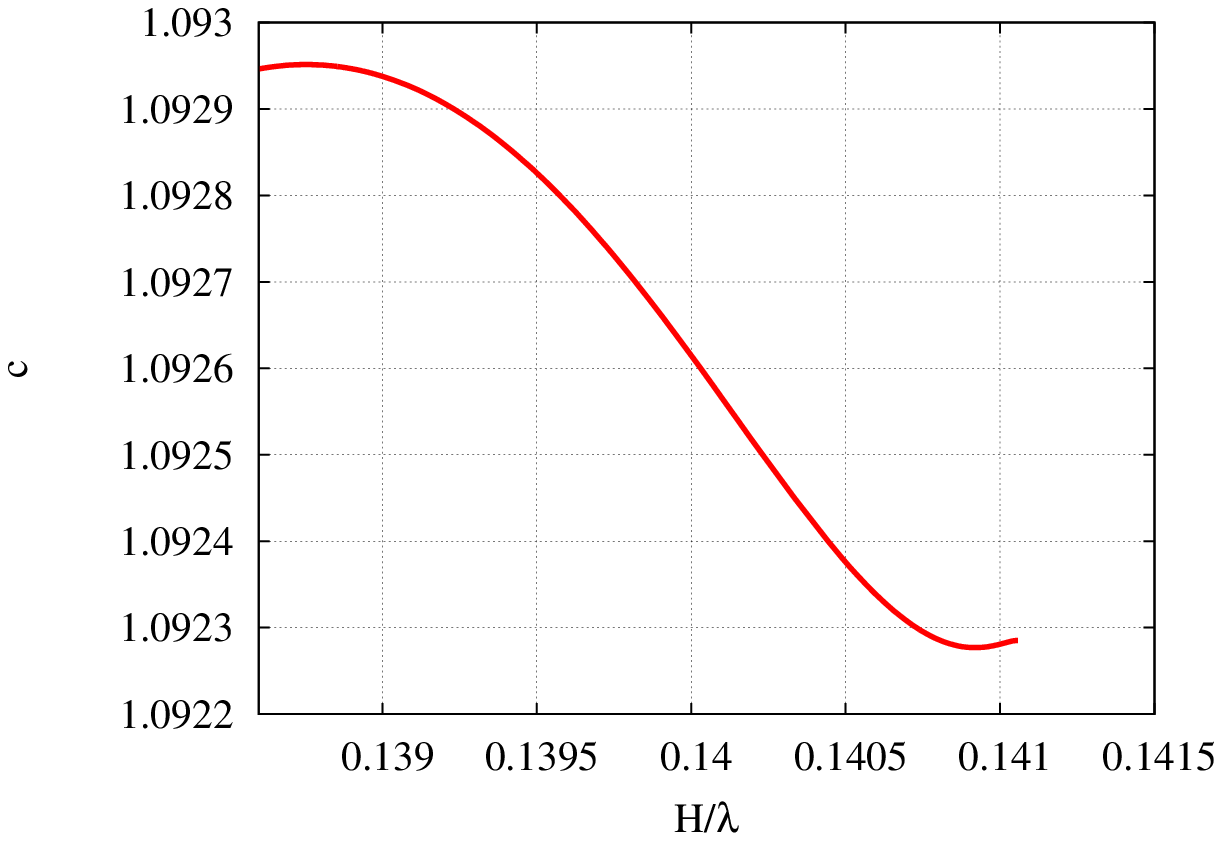}\\
\includegraphics[width=2.65in]{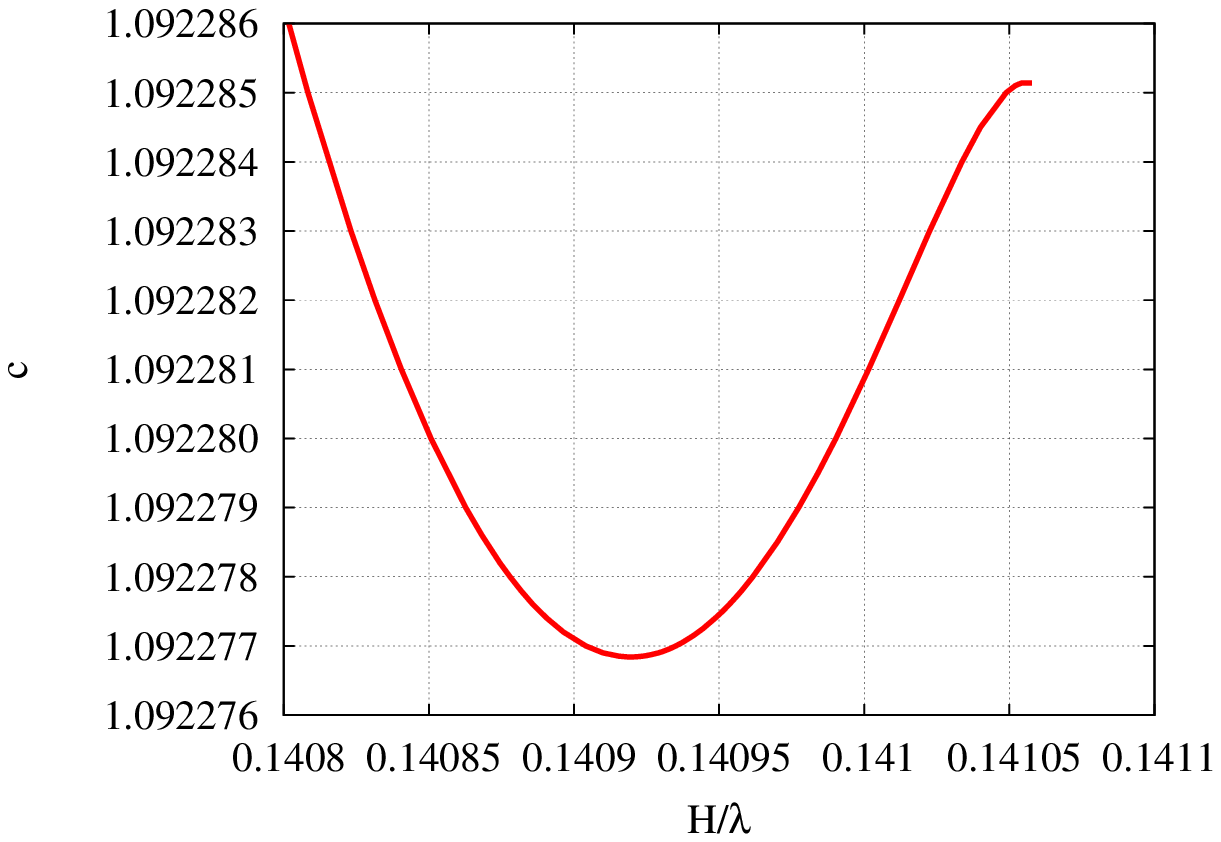}
\includegraphics[width=2.65in]{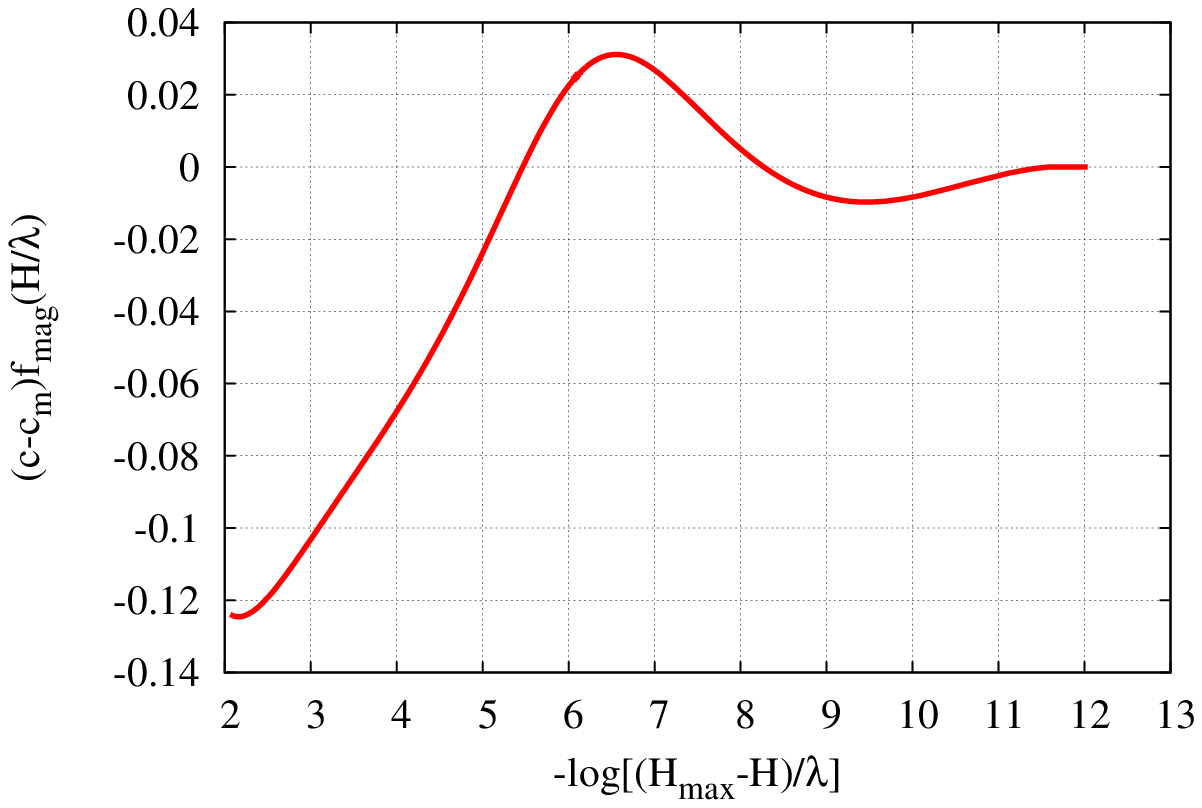}
\caption{\label{oscillations}
Oscillations of dimensionless velocity of Stokes wave propagation as function of steepness obtained from simulations. First three plots from left to right and from top to bottom have increasing zoom both in vertical and horizontal axes to focus on oscillations. In the lower right corner a plot is scaled by a magnification function $f_{mag}(H/\lambda) = 1/(30(H_{max}-H)/\lambda)^{1.15} + 1$ to show all simulation data in a single graph while stressing obtained oscillations.
}
\end{figure}

Results of multiple simulations of Stokes wave are shown in Figure \ref{oscillations}, where the wave velocity $c$ is shown as a function of the dimensionless wave height $H/\lambda$. This function is nonmonotonic which is in agreement with previous simulations (e.g.~\citet{SchwartzJFM1974,Williams1981,WilliamsBook1985}) and theoretical analysis (\citet{Longuet-HigginsFoxJFM1977,Longuet-HigginsFoxJFM1978}) which predicted an infinite number of oscillations.

We were able to resolve with quadruple precision two oscillations
(two maxima and two minima) of the propagation velocity as a
function of $H/\lambda$. These oscillations
represent a challenge for simulation, because propagation velocity
is the only parameter in the equation~\eqref{stokes_wave2}. Then it is impossible even to go over  the first maximum by changing continuously velocity of
propagation $c$. This is
because after the maximum is reached, we has to start decreasing
the parameter $c.$ But decreasing of $c$ causes iterations  to converge to the less steep
solution on the left from the maximum (which we already obtained on previous steps),
instead of steeper solutions to the right from the maximum.

\begin{figure}
\centering
\includegraphics[width=5in]{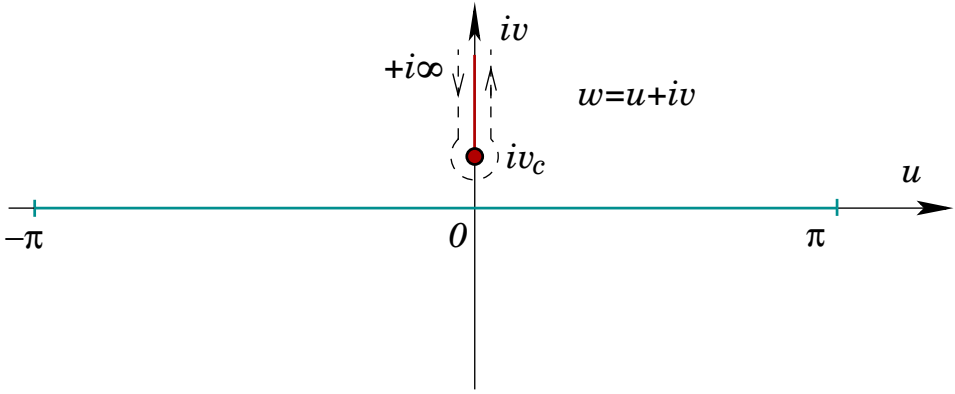}
\caption{\label{contour_map} Schematic of contour in the $\mathbb{C}^+$ which allows to determine distance $v_c$
from the branch cut to the real axis.}
\end{figure}

In order to resolve this issue we used the following approach. 
Assume that the singularity of $\tilde z$ closest to real axis in $w$ complex plane is the branch point  %
\begin{align} \label{ztildebranchpoint}
\tilde z\simeq c_1(w-\I v_c)^\beta
\end{align}
 for $w\to \I v_c$, where $c_1$ is the complex constant, $v_c>0$ and $\beta$ are real constants.  By the periodicity in $u$, similar branch points are located at $w=\I v_c+2\upi n, \, n=\pm1,\pm 2,\ldots$ (recall that that we already switched to the dimensionless coordinates). We expand $\tilde z(u)$ into Fourier series $\tilde z(u)=\sum\limits_{-\infty}^{k=0} \hat {\tilde  z}_k\exp(\I ku)$, where
\begin{align} \label{ztildefourier}
\hat {\tilde  z}_k=\frac{1}{2\upi}\int\limits_{-\upi}^{\upi} \tilde z(u)
\E^{-\I ku}\D u
\end{align}
are Fourier coefficients and the sum is taken over nonpositive
integer values of $k$ which ensures both $2\upi$-periodicity of
$\tilde z(u)$ and analyticity of $\tilde z(w)$ in $\mathbb{C}^-$.  We
evaluate \eqref{ztildefourier} in the limit $k\to -\infty$ by moving
the integration contour from the line $-\upi<u<\upi$ into $\mathbb{C}^+$  until
it hits the lowest branch point \eqref{ztildebranchpoint} so it goes
around branch point and continues straight upwards about both sides
of the corresponding branch cut as shown by the dashed line in right
panel of Fig.~\ref{contour_map}.
Here we assume that branch cut
is a straight line connecting $w=\I  v_c$ and $+\I \infty$. Then the
asymptotic of $|\hat {\tilde  z}_k|$ is given by
\begin{align} \label{zkasymptotic}
|\hat {\tilde  z}_k|\propto |k|^{-1-\beta}e^{-|k|v_c}, \quad k\to
-\infty.
\end{align}

This approach was used in our previous work~\citep{DLK2013} to evaluate distance $v_c$ of the lowest singularity
to the real line. Now our key idea is to push artificially the singularity  $w=i v_c$ toward
the real line, thus increasing  $H/\lambda$.  It follows
from expression~\eqref{zkasymptotic} that to decrease $v_c$ we can multiply Fourier coefficients of the previously obtained Stokes wave solution $\hat {\tilde  z}_k$ by $\exp(\alpha k)$,
where  the numerical parameter $\alpha$ is chosen such that $0<\alpha\ll v_c$. The result of this multiplication  $\hat {\tilde  z}_k\exp(\alpha k)$ is not a Stokes wave solution anymore,
but it has higher steepness and not very distinct from the Stokes wave solution if $\alpha$ is small enough. After that modification  we slightly decrease $c$ from previous value and allow iterations of Section \ref{sec:NewtonCG} to converge starting from   $\hat {\tilde  z}_k\exp(\alpha k)$ as zero iteration. As we expected, iterations then converge  to the solution on the right from the maximum. This procedure
allowed us to resolve both maxima and one nontrivial minimum of $c$ as a function of $H/\lambda$ as summarized in Fig.~\ref{oscillations}.

\subsection{Recovering $v_c$ from the Fourier spectrum of Stokes wave}
\label{sec:RecoveringvcFourierspectrum}

To obtain the location of the branch point $w=iv_c$ with good precision one has to go beyond the leading order asymptotic \e{zkasymptotic}.
Next order corrections to the integral \e{ztildefourier} for $\beta=1/2$ have the following form
\begin{align} \label{zkasymptotic2}
|\hat {\tilde  z}_k|\simeq \left (c_1|k|^{-3/2}+c_2|k|^{-5/2}+c_3|k|^{-7/2}+c_4|k|^{-9/2}+\ldots\right )e^{-|k|v_c}, \quad k\to
-\infty,
\end{align}
where we took into account the expansion of $\tilde z(w)$ in half-integer powers $(w-\I v_c)^{1/2+n}, \ n=0,1,2,\ldots$ beyond the leading order term \e{ztildebranchpoint}.

The numerically obtained spectrum $|\hat {\tilde  z}_k|$ of Stokes
wave was fitted to the expansion \e{zkasymptotic2} in order to
recover $v_c$ and coefficients $c_1, \ c_2, \ c_3, \ldots$. The
highest accuracy in recovering $v_c$ was achieved when the middle of
spectrum $k\sim k_{max}/2$ was used for that fit, there
$k_{max}=M/2$ is the highest Fourier harmonic used in simulations.
$k_{max}/2$ represent a compromise between the highest desired
values of $k$ to be as close as possible to asymptotic regime $k\to
\infty$ and the loss of numerical precision for $k\to k_{max}$. We
estimated the accuracy of the fit by varying values of $k$ used for
fitting as well as changing the number of terms in the expansion
\e{zkasymptotic2}. Typically we used 4 terms in \e{zkasymptotic2}.
Section \ref{sec:Findingchic} discusses the comparison of the
accuracy of the obtained results with the other methods we used to
find $v_c$.

\subsection{Highest wave obtained}
We calculated $\tilde  z(u)$ with high accuracy for different values
of  $H/\lambda$ using computations in quad precision (32 digits). Such high precision is necessary to reveal
the structure of singularities in $\mathbb{C}^+$.
\begin{figure}
\centering
\includegraphics[width=3.2in]{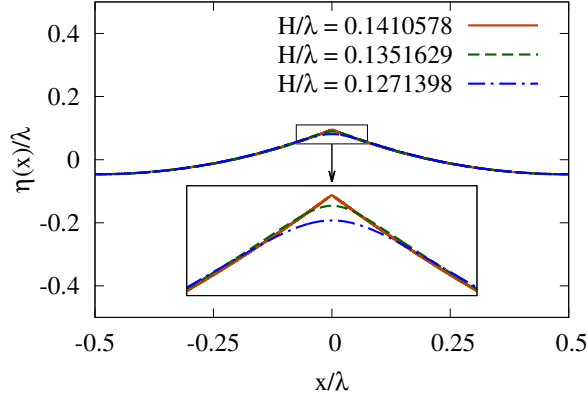}
\caption{\label{many_waves}Stokes wave with
$c=1.082$ (blue dash-dotted line), $c=1.091$ (green dashed line) and
$c=1.0922851405$ (dark orange solid line). Corresponding values of
$H/\lambda$ are given in the legend. Inset shows zoom-in into small
values of $x/\lambda$ near a wave crest. }
\end{figure}
Fig. \ref{many_waves}
shows  spatial profiles of Stokes waves for several values of
$H/\lambda$ in physical variables $(x,y).$ The Stokes wave quickly
approaches the profile of limiting wave except a small neighborhood
of the crest.

As it was shown in the previous paper~\citet{DLK2013}, tails of the
spectra have asymptotic behavior corresponding to $\beta=1/2$ in
\eqref{zkasymptotic}, which means that we have square root branch
cut singularity in $\mathbb{C}^+$ all the time. This is consistent
with theoretical predictions by
\citet{MalcolmGrantJFM1973LimitingStokes} and \citet{TanveerProcRoySoc1991}.


The number of Fourier modes $M\equiv 2k_{max}$ which we used in Fast Fourier Transform
(in simulations we expand $y(u)$ in cosine Fourier series to speed
up simulations and to be memory efficient) for each value
$H/\lambda$ increases quickly with the increase of $H$ as $v_c$
decreases. E.g., for $H/\lambda=0.0994457$ it was more than enough
to use 256 modes while for the largest wave height
\begin{align} \label{Hmaxnum}
H_{max}^{num}/\lambda=
                      0.141057778854883208164928602256956
\end{align}
achieved in simulations we used $M=2^{27}\approx 134\times10^{6}$ modes. Due to such high number of modes,
 the precision of value \eqref{Hmaxnum}
 decreases by round-off errors in approximately $M^{1/2}$ times, i.e. in $\sim 4$ digits. This extreme case has $c=1.0922851405$ and
  $v_c = 5.93824419892803271779  \times10^{-7}.$ These numbers are the moderate extension of our previous work ~\citet{DLK2013} by pushing down a lowest value of $v_c$ more than twice.  Further decrease of the numerical values of $v_c$ can be achieved by both subtracting the leading order singularity \e{ztildebranchpoint} from the numerical solution and using  the nonuniform
  numerical grid in $u$ which concentrates near $u=0. $ These numerical approaches are however beyond the scope of this papers.

Before our work \citet{DLK2013},  the numerical estimates of $H_{max}$ were found
by~\citet{WilliamsBook1985} as $H^{Williams}_{max}/\lambda=
0.141063$  and \citet{GandzhaLukomskyProcRoySocLond2007}  $H^{GL}_{max}/\lambda=
0.14106348398.$  The other commonly used
but less precise estimate is $H^{Schwartz}_{max}/\lambda=
0.1412$~\citep{SchwartzJFM1974}. It was shown in~\citet{DLK2013}  that numerical values of $v_c$ in
the limit  $(H_{max}-H)/\lambda\ll 0$ was fitted to  the scaling law
\eqref{vcscalinglaw} with
\begin{align} \label{Hmaxnew}
H_{max}/\lambda=0.1410633\pm 4\cdot10^{-7}.
\end{align}
The mean-square error for $\delta$ in \eqref{vcscalinglaw} is
$\simeq 0.04$ which offers the exact value $\delta=3/2$ as a
probable candidate for  \eqref{vcscalinglaw}. The estimate
\eqref{Hmaxnew} suggests that the previous estimate
$H^{Williams}_{max}$ is more accurate than $H^{Schwartz}_{max}$.  Also $H^{GL}_{max}$
is within the accuracy of the estimate
\eqref{Hmaxnew}. However,  $H^{GL}_{max}$
 is obtained in Ref.  \citet{GandzhaLukomskyProcRoySocLond2007}   from  the Michell's expansion~\citep{Michell1893}  of the limiting Stokes wave which ignores the expansion in powers of the irrational number $\mu=1.46934574\ldots$ Existence of that expansion beyond the Stokes power law $u^{2/3}$  was established by \citet{MalcolmGrantJFM1973LimitingStokes}.  Lack of resolving that expansion suggests that $H^{GL}_{max}$
 does not have a well controlled accuracy. In contrast, our numerical results are based on  Fourier series for non-limiting Stokes wave which has well-controlled precision.   The
difference between   \eqref{Hmaxnew}  and the new  lower boundary estimate
\eqref{Hmaxnum} of the largest $H$  is $\simeq0.004\%$.

\section{Pad\'e approximation of Stokes wave}
\label{sec:PadeApproximationStokesWave}

\subsection{Additional conformal transformation and spectral convergence of Pad\'e approximation}

\begin{figure}
\centering
\includegraphics[width=5in]{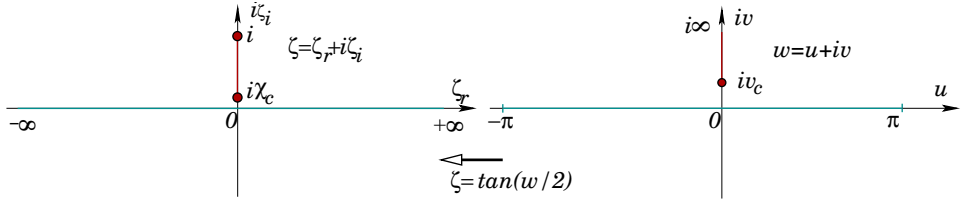}
\caption{\label{second_map} Schematic of a second conformal map
between the periodic domain in $w$-plane (right panel) into
$\zeta=\tan(w/2)$-plane. Another useful property of this map
is representation of $2\upi$-periodic branch cut from $\I v_c$ to $\I\infty$ as a finite
length cut from $\chi_c=\tan(\I v_c/2)$ to $\I$.}
\end{figure}

To analyze the structure of singularities of Stokes wave we perform an additional conformal transformation between the complex plane  $w=u+\I v$  and  the complex plane for the new variable
\begin{equation} \label{zetadef}
\zeta=\tan\left (\frac{w}{2}\right ).
\end{equation}
Equation \e{zetadef} maps the strip $-\upi <Re(w)<\upi$ into  the
complex $\zeta$ plane. In particular, the line segment
$-\upi<w<\upi$ of the real line $w=u$   maps into the real line
$(-\infty,\infty)$ in the complex plane $\zeta$ as shown in
Fig.~\ref{second_map}. Vertical half-lines $w=\pm\upi+\I v, \
0<v<\infty$ are mapped into a branch cut   $\I<\zeta<\I\infty.$ In a
similar way, vertical half-lines $w=\pm\upi+\I v, \ -\infty<v<0$ are
mapped into a branch cut   $-\I\infty<\zeta<\I.$ However,
$2\upi-$periodicity of $\tilde z(w)$  \e{xtildeperiodic} allows to
ignore these two branch cuts because  $\tilde z(w)$ is continuous
across them.   Complex infinities $w=\pm  \I\infty $ are mapped into
$\zeta=\pm \I$. An unbounded interval $[\I v_c,\I\infty),$ $v_c>0$
is mapped into a finite interval $[\I\chi_c,\I)$ with
\begin{align} \label{chicdef}
\chi_c=\tanh{\frac{ v_c}{2}}.
\end{align}
 The mapping  \e{zetadef} is different from the commonly used (see e.g. \cite{SchwartzJFM1974,Williams1981,TanveerProcRoySoc1991}) mapping $\zeta =\exp{(-\I w)}$ (maps the strip  $-\upi \le Re(w)<\upi$   into the unit circle). The advantage of using the mapping  \e{zetadef} is the compactness of the interval $(\I\chi_c,\I)$ as mapped from the infinite interval $(\I v_c,\I\infty)$. In contrast, the mapping to the circle leaves the interval $(\I v_c,\I\infty)$ infinite in $\zeta$ plane.

We use Alpert-Greengard-Hagstrom (AGH)  algorithm
\citep{AGH2000,LauClassQuantumGrav2004} to approximate the Stokes wave  $\tilde z(\zeta)$ at the real line $\Real(\zeta)=\zeta$  by a set a poles  in the complex $\zeta$ plane. Approximation by a set of poles is a particular case of  Pad\'e
approximation by  rational functions $\frac{P(\zeta)}{Q(\zeta)}$, where $P(\zeta)$ and $Q(\zeta)$ are polynomials. Zeros of $Q(\zeta)$ give the location of poles.  Looking at complex values of $\zeta$ in the rational function   $\frac{P(\zeta)}{Q(\zeta)}$
provides the analytical continuation of   $\tilde z(\zeta)$ into the complex $\zeta$ plane.
Usually   Pad\'e
approximation is numerically unstable because of the pairs of spurious zeros and poles  appear in finite precision arithmetics. These doublets correspond to positions of zeros of    
 $P(\zeta)$ and $Q(\zeta)$ which are nearly cancel each other.  In our practical realizations, AGH algorithm avoids the
numerical instability of the Pad\'e approximation until the number
of poles $N$ increases to reach the accuracy corresponding to the
round-off error in the numerical approximation of $\tilde z(u).$  If
the analytical continuation of $\tilde z(u)$ into $w\in \mathbb{C}$
has a branch cut, the AGH algorithm places poles along the branch
cut. AGH algorithm is outlined in Appendix~\ref{sec:AGHalgorithm}.

\begin{figure}
\includegraphics[width=0.8\textwidth]{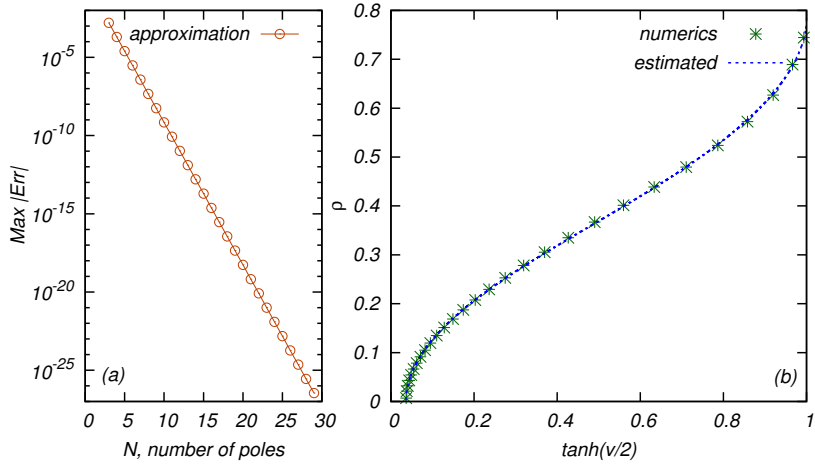}
\caption{\label{AP1:err_density}(a) An exponential decay of error in
Pad\'e approximation of Stokes for
$H/\lambda=0.125510247666212033511898125908053$ as a function of the
number of poles $N$. (b) The density $\rho(\chi)$   on the branch
cut sampled at $\zeta_k = \I\chi_k\equiv
\I\tanh{\left(\frac{v_k}{2}\right)}$, $k=1,,\ldots,N,$ obtained from
the Pad\'e approximation of Stokes wave  from(a) with $N=29$ (green
stars). Blue dotted line is the estimated profile of $\rho(\chi)$
for the same Stokes wave in the continuous limit of $N\to\infty$.}
\end{figure}

\begin{figure}
\includegraphics[width=0.8\textwidth]{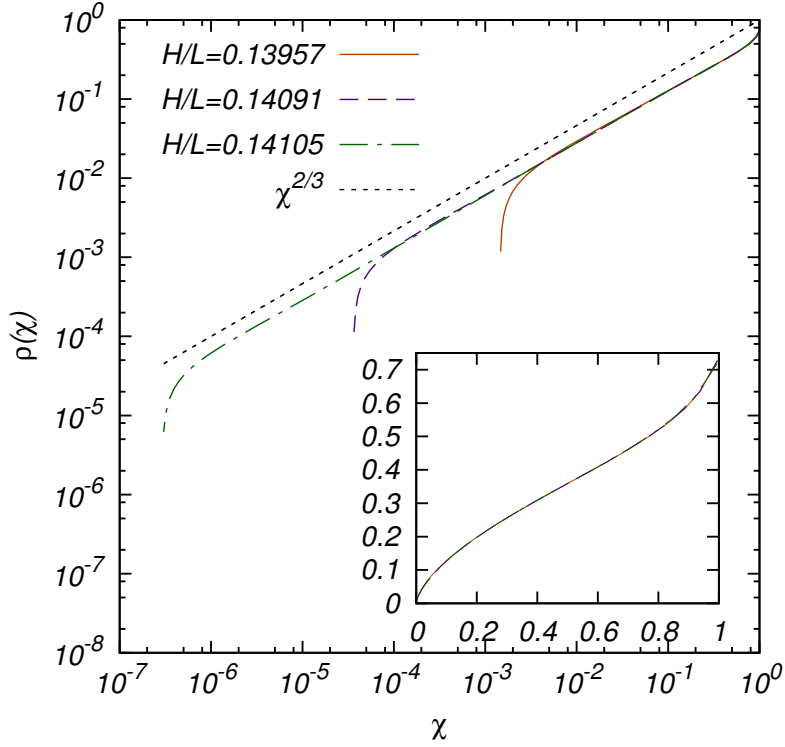}
\caption{\label{fig:logdensity} The density $\rho(\chi)$ for three different Stokes waves in log-log scale. A straight dashed line shows $\chi^{2/3}$ scaling law which corresponds to the limiting Stokes wave. Insert shows  $\rho(\chi)$ in linear scale for the same three Stokes wave which are visually almost indistinguishable. }
\end{figure}

We applied AGH algorithm for    $\tilde z(\zeta)$  at the real line $\Real(\zeta)=\zeta,$ where    $\tilde z(\zeta)$  is obtained from   simulations described in Section \ref{sec:NumericalStokes}. Increasing $N$ we observed the exponential convergence of Pad\'e
approximation $z(\zeta)_{pade}$ to    $\tilde z(\zeta)$ as
\begin{equation} \label{errinf}
err_\infty\propto e^{-p(v_c)N},
\end{equation}
where $err_\infty\equiv\max\limits_{-\infty<\zeta<\infty} |\tilde
z(\zeta)-   \tilde z(\zeta)_{pade}|$ is the error in infinity (maximum) norm.
An example of the exponential convergence is  shown in Figure
\ref{AP1:err_density}a for a particular Stokes wave. Here $p(v_c)$
is the function of $v_c$ but is independent on $N$. We found that
with high precision
\begin{equation} \label{errscaling}
p(v_c)\propto v_c^{1/6}.
\end{equation}
 AGH algorithm is looking for poles in the
entire complex plane $\zeta$.   All the encountered poles for Stokes wave were
found on the interval of imaginary axis along the interval
$[\I\chi_c, \I)$, where $\chi_c$ is determined numerically as in Section \ref{sec:NumericalStokes}.

Equations \e{errinf} and \e{errscaling} demonstrate excellent performance of Pad\'e approximation. E.g., decreasing $v_c$ by six order required in our simulations only 10-fold increase of $N$ as detailed in Appendix C. It suggests that numerical method which solves Stokes wave equation \e{stokes_wave2} directly in terms of Pad\'e approximants might be  superior to  Fourier methods including   numerical approaches mentioned in Section \ref{sec:RecoveringvcFourierspectrum}. This topic is
however beyond the scope of this paper.

It is rather straightforward to distinguish in AGH algorithm poles
from branch cuts. If both poles and branch cuts would be present in
$\tilde z(\zeta)$ then increasing $N$ one observes that some poles
of Pad\'e approximation are not moving and their complex residues
remain approximately the same. These correspond to poles of $\tilde
z(\zeta).$   Such behavior occurs for test problems when we
artificially added extra poles to      $\tilde z(\zeta)$.   Other
poles of Pad\'e approximation are moving with the increase of $N$
and their complex residues are  changing.    These poles mark the
spatial location of branch cuts of       $\tilde z(\zeta)$. The
density of poles along each branch cut is increasing   with the
increase of $N$. If the jump of       $\tilde z(\zeta)$ at branch
cut is continuous along it then we expect to see the convergence of
density of poles with the increase of $N$.  All this is valid until
$err_\infty$ decreases down to the level of round-off error at which
$\tilde z(\zeta)$ was determined. Further increase of $N$ would
result in the appearance of spurious poles at random positions of
$\zeta$ plane with the magnitudes of complex residues at the level of
round off error ($\sim 10^{-32}$ for $z(\zeta)$ found with quad
precision in Section \ref{sec:NumericalStokes}).

Using       $\tilde z(\zeta)$ obtained by the method of  Section
\ref{sec:NumericalStokes}, we found a single branch cut $[\I\chi_c,
\I)$ but no poles in Stokes wave. It means that in complex $w$ plane
we have one branch cut per spatial period $2\upi$ located at $(2\upi
n+\I v_c,2\upi n+\I \infty)$, $n\in\mathbb{N}.$

We parametrize that branch cut as follows
\begin{equation}
\label{branchcutdensity} \tilde z(\zeta) = \I y_0 +
\int\limits_{\chi_c}^{1} \dfrac{\rho(\chi')\D \chi'}{\zeta - \I\chi'}
,
\end{equation}
where $\rho(\chi)$ is the density along branch cut. That density is
related to  the jump of $\tilde z(\zeta)$ at branch cut as explained
in Section \ref{sec:jumpbranchcut}. The constant $ y_0$ is
determined by the value of $\tilde z(\zeta)|_{\zeta=\infty}=\tilde
z(w)|_{w=\upi}.$ This constant has a zero imaginary part, $\text{Im}(y_0)=0,$ because $\tilde
x(w)|_{w=\upi/2}=0$ as given by the equation \eqref{xtildeperiodic}.

The Pad\'e approximation represents equation  \e{branchcutdensity}
as follows
\begin{equation}
\label{approx_cut} \tilde z(\zeta) = \I y_0 + \int\limits_{\chi_c}^{1}
\dfrac{\rho(\chi')\D \chi'}{\zeta - \I \chi'} \simeq \I  y_0+
\sum\limits_{n = 1}^{N} \dfrac{\gamma_n}{\zeta - \I \chi_n},
\end{equation}
where the numerical values of the pole positions $\chi_n$ and the
complex residues $\gamma_n$ ($n=1,\ldots, N $) are obtained from AGH
algorithm.

\subsection{Recovering jump along branch cut}

We recover $\rho(\chi)$ from equation \e{approx_cut} as follows. Assume that we approximate the integral in equation \e{branchcutdensity} by the trapezoidal rule
\begin{align} \label{trapezoid}
\int\limits_{\chi_c}^{1} \dfrac{\rho(\chi')\D \chi'}{\zeta - \I \chi'}
\simeq&\frac{\chi_2-\chi_{1}}{2} \dfrac{\rho_1}{\zeta -
\I \chi_1}\nonumber\\&+ \sum\limits_{n =2}^{N-1}
\frac{\chi_{n+1}-\chi_{ n-1}}{2} \dfrac{\rho_n}{\zeta -
\I \chi_n}+\frac{\chi_N-\chi_{N-1}}{2} \dfrac{\rho_N}{\zeta -
\I \chi_N}.
\end{align}
A comparison of equations \e{approx_cut} and \e{trapezoid} suggests the approximation $\rho_{n,N}$ of the density $\rho(\chi_n)$ on the discrete grid $\chi_n$, $n=1,\ldots, N$  as follows
\begin{subequations} \label{rhok}
\begin{align}
&\rho(\chi_n)\simeq\rho_{n,N}=\frac{2\gamma_n}{\chi_{n+1}-\chi_{n-1}} \ \text{for} \ n=2,\ldots, N-1, \label{rhoka}\\
&  \rho(\chi_1)\simeq\rho_{1,N}=\frac{2\gamma_1}{\chi_{2}-\chi_1};  \ \rho(\chi_N)\simeq\rho_{N,N}=\frac{2\gamma_N}{\chi_{N}-\chi_{N-1}}. \label{rhokb}
\end{align}
\end{subequations}

\begin{figure}
\begin{center}
\includegraphics[width=5in]{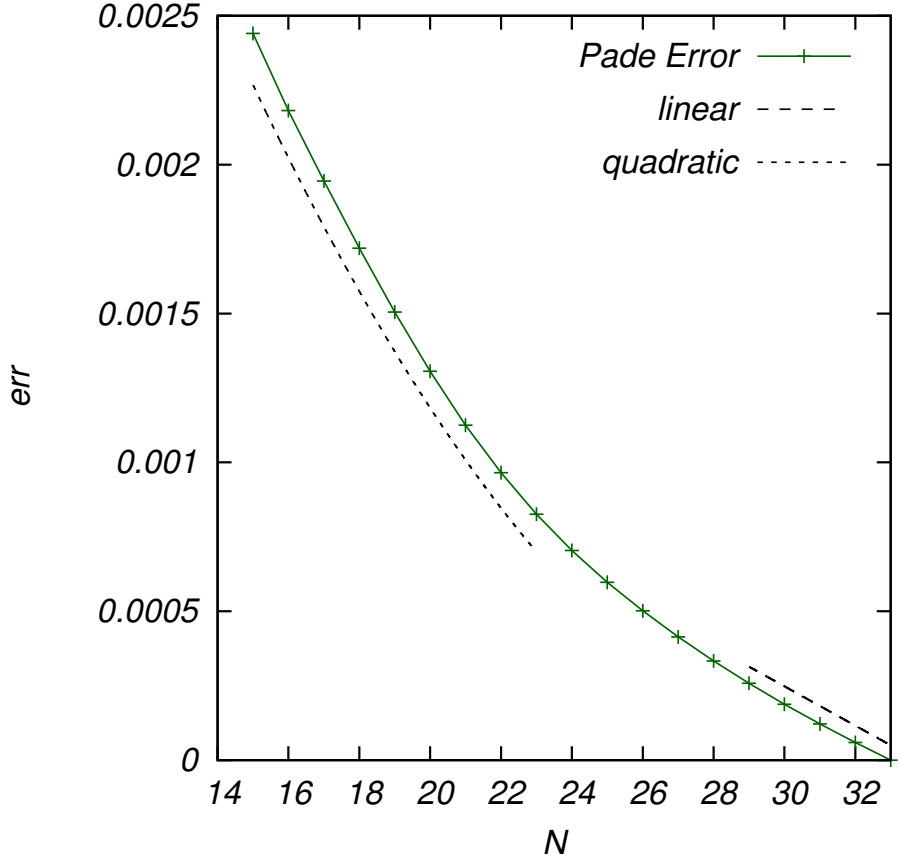}
\end{center}
\caption{\label{fig:padeerr} Error between Pad\'e approximation with $N$ poles and the continuous limit for $\rho(\chi)$ for Stokes wave with $H/\lambda=0.125510247666212033511898125908053.$ It is seen that the error $\propto\frac{1}{N^2}$ for large $N$. To calculate that error we use a spline interpolation for   $\rho^2_{n,N}$ with $N=N_{max}=33  $ to construct the approximation of the continuous limit of the square of the density, $\rho^2_{continuous}(\chi).$  After that the error is defined as $\text{err}\equiv\left (\sum^{N-1}_{n=2}[\rho^2_{n,N}-\rho^2_{continuous}(\chi_n)]^2\right )^{1/2}/(N-2)$ for each $N$, where $\rho_{n,N} $ is given by equation \e{rhok}.}
\end{figure}

 A convergence of $\rho_{n,N}$ to the continuous limit $\rho(\chi_n)$ as $N$ increases is quadratic with the error scaling $\propto \frac{1}{N^2} $ for $\chi$ away
 from boundaries $\chi=\chi_c$ and $\chi=1.$ Near these boundaries we cannot apply the trapezoidal rule and have to resort to less accurate estimates given by the
 equation \e{rhokb}.   Figure \ref{fig:padeerr} demonstrates this   $\propto\frac{1}{N^2} $ convergence of the Pa\'de approximation to the continuous limit. Figure \ref{AP1:err_density}b shows the particular example of $\rho(\chi)$ (shown by solid line solid line) compared with  $\rho_{n,N}$  (shown by stars) for $N=29$. We believe that the convergence of $\rho_n$ to the continuous value $\rho(\chi)$ as $N\to \infty$ and absence of other poles outside of $[\I\chi_c,\I]$ provide a
 numerical proof that the only singularity of $\tilde z(\zeta)$ are the branch points $\I\chi_c$ and $\I$ connected by the branch cut
 $\zeta\in[\I\chi_c,\I]. $

At $\chi=\chi_c$ the function $\rho(\chi)$ has a square root singularity as given below by equation \e{halfser}. This singularity additionally reduces the accuracy of the approximation  \e{rhokb} for $\rho_{1,N}$ which is based  on Taylor series.
 To significantly improve numerical accuracy of $\rho_{1,N}$ we assume that $\rho$ has  the following square root dependence in the vicinity of $\chi_c$:
 \begin{align}\label{Achic}
 \rho_{approx}(\chi) = A\sqrt{\chi-\chi_c}.
 \end{align}
 Here the values of the parameters $A$ and $\chi_c$ are  determined from two interior points $(\chi_2, \rho_{2,N} )$ and $(\chi_3,\rho_{3,N} )$ found via the trapezoid rule~\eqref{rhoka}. We assume that $\rho_{approx}(\chi_2)=\rho_{2,N}$ and  $\rho_{approx}(\chi_3)=\rho_{3,N}$  which gives
that \begin{align} \label{Achican}
 A = \left(\dfrac{\rho_{3,N} ^2 - \rho_{2,N} ^2}{\chi_3 - \chi_2}\right)^{1/2}, \qquad
 \chi_c = \dfrac{\rho_{3,N} ^2\chi_2 - \rho_{2,N} ^2\chi_3}{\rho_{3,N} ^2-\rho_{2,N} ^2}.
 \end{align}
 Using equations \e{Achic} and \e{Achican} for $\chi=\chi_1$ we obtain the numerically accurate approximation that
\begin{align}\label{rho1an}
 \rho_{1,N} = \left ({\dfrac{(\chi_3-\chi_1)\rho_{2,N}^2 - (\chi_2-\chi_1)\rho_{3,N}^2 }{\chi_3-\chi_2}}\right )^{1/2},
 \end{align}
 where  $\rho_{2,N} $ and $\rho_{3,N}$ are given by equation~\eqref{rhoka}.

At  $\chi=1$  the function $\rho(\chi)$ also has singularity and respectively numerical value of $\rho_{N,N}$ from  ~\eqref{rhokb}  is not very accurate. To improve that accuracy we use that $\rho(1)=1$ as found in Part II. Then using the trapezoidal rule we obtain much more accurate expression that
\begin{align} \label{rhoNN}
 \rho(\chi_N)\simeq\rho_{N,N}=\frac{2\gamma_N}{1-\chi_{N-1}}.
\end{align}

Figure \ref{fig:logdensity} shows the density $\rho(\chi)$ for three different Stokes waves in log-log
 scaling. It is also seen that inside the branch cut and for small $\chi_c\ll 1$,   the density
 $\rho(\chi)$ scales as $\chi^{2/3}$  which corresponds to the
 limiting Stokes wave. A deviation from that scaling occurs near
 $\chi=\chi_c$ and $\chi=1.$

Classical Markov's theorem \citep{MarkovActaMath1895} proves pointwise convergence of the diagonal Pad\'e approximants $[N/N]_f$ of the function $f$ of the type  \e{branchcutdensity} with $\rho(\chi)\ge 0$ in the  limit $N\to \infty $ for  $\zeta\in\C\setminus[\I\chi_c,\I].$  Here the diagonal Pad\'e  approximation $[N/N]_f$ of the function $f$ means that both polynomials  $P(\zeta)$ and $Q(\zeta)$ has the same order $N$ which is natural for the discretization \e{approx_cut}. More general  Pad\'e  approximants of the function $f$ are   $[N/M]_f,$ where $N$ and $M$ are  the orders of the polynomials  $P(\zeta)$ and $Q(\zeta), $ respectively.   Theorem of \citet{deMontessusdeBallore1902} ensures  pointwise convergence of $[N/M]_f\to f$ for $N\to \infty$ with fixed $M$ in the disk $|\zeta|<R$ if $f$ the meromorphic function in that disk with exactly $M$ poles (counted according to their multiplicity).  However, the diagonal Pad\'e approximations of the meromorphic function $f$ generally fails to provide uniform convergence with the known counterexamples given by \citet{Buslaev2001,Lubinsky2003}.   \citet{Nuttall1970} showed that instead the diagonal Pad\'e approximants of meromorphic function for $N\to \infty$ have a weaker convergence in logarithmic capacity which allows the lack of pointwise convergence along exceptional sets.    Gonchar \citep{Gonchar1973,Gonchar1975} extended Markov's theorem  on the pointwise convergence of the diagonal  Pad\'e approximants to the functions $f+r$, where $f$ is the function of the type  \e{branchcutdensity} with $\rho(\chi)>0$ almost everywhere in $\chi\in[\chi_c,1]$ and $r$ is the meromorphic function away from branch cut and has no poles at branch cut.  Convergence in logarithmic capacity of the diagonal Pad\'e approximants of the analytic  function $f(\zeta)$ with a finite number of branch points  (this is a more general type than the type  \e{branchcutdensity} because these branch points can be located away from a single line)    was proven  by
\citet{StahlComplexVariablesTheoryApplI1985,StahlComplexVariablesTheoryApplII1985,StahlJApproxTheory1997}. That convergence occurs away  from certain sets of $\C$ (in some cases these sets are simple arcs). See also Ref. \citet{AptekarevBuslaevMartinezinkelshteinSuetinRussianMathSurveys2011} for the recent review.  All these results were obtained for Pad\'e approximants based on the Taylor series at a single point in $\C.$   Thus these  results  do not directly apply to AGH algorithm which is based on least squares approximation at multiple points of $\C.$  AGH algorithm is also distinct from  multipoint Pad\'e approximation \citep{Saff1972,BakerGravesMorrisBook1996}, where the Taylor series is interpolated at multiple points  in contrast to least squares in AGH algorithm.  Pad\'e approximants were also constructed    based on least squares in  Ref.
\citet{GonnetPachonTrefethenElectrTransNumAnal2011} were it was conjectured   that least squares-type algorithms might ensure pointwise convergence. That conjecture is consistent with our simulations.

\subsection{Finding a numerical value of a location of branch point $\zeta=i\chi_c$ }
\label{sec:Findingchic}

There are  different ways to find the location of branch point $\zeta=\I \chi_c$ from simulations. First way is based on the decay of Fourier spectrum of          $\tilde z(\zeta)$ for $n\gg 1$ and is described in Section  \ref{sec:RecoveringvcFourierspectrum}. Second way is to  find $\rho(\chi_n)$, $n=1,\ldots, N$ and then determine the point $\rho(\chi_c)=0$ by the polynomial extrapolation of  $\rho(\chi_n)$.   First and second ways provide comparable numerical accuracy in our simulations (typically the relative error in $\chi_c$ is $\sim 10^{-4}$).

We found however, that  better accuracy is achieved in the third way as follows. Consider the formal series
\begin{equation} \label{halfser}
\tilde z_{ser}=\sum\limits_{j=0}^\infty ie^{\I
j\upi/4}a_{j}(\zeta-\I \chi_c)^{j/2}
\end{equation}in the neighborhood of the branch point $\zeta=\I \chi_c.$
The term $ie^{\I j\upi/4}$ in front of the coefficients $a_j$ is
chosen for convenience to ensure that coefficients $a_j$ take real
values. The radius of convergence of that series is $2\chi_c$ as
discussed in Part II. Taking $M=10-20$ terms in that series one can
use the nonlinear fit to determine the unknowns $\chi_c$ and $a_j.$
Typically we use $N_j=30-40$ points   $(u_n,\tilde z(u_n)) $ such
that all values $u_n$ are inside the disk of convergence $|u_n -\I
\chi_c|<2\chi_c$ of the series \e{halfser}. Here  values of $\tilde
z(u_n)$  are taken from simulations of  Section
\ref{sec:NumericalStokes} with $u_n$ being the numerical grid points
closest to $u=0.$    The accuracy of the nonlinear fit  is typically
$\sim 10^{-10}$ as estimated by varying $M$ and $N_M.$ In Part II we
provide much more accurate way of calculating $\chi_c$ which is
based on the compatibility of the series  \e{halfser}  with the
equation \e{stokes_wave2} of  Stokes wave. In contrast, the above
three  methods use numerical values of  $(u_n,\tilde z(u_n))$
obtained as described in Section \ref{sec:NumericalStokes} and do
not use the equation \e{stokes_wave2} directly.

\section{Stokes wave as an integral over jump at branch cut and the expansion of density $\rho$ near a branch point }
\label{sec:Stokeswaveasintegral}

\subsection{Jump at  branch cut}
\label{sec:jumpbranchcut}

Sokhotski–-Plemelj theorem  (see e.g. \cite{Gakhov1966,PolyaninManzhirov2008}) applied to the equation \e{branchcutdensity} gives

\begin{equation}
\label{branchcutdensitySokhotski} \tilde z(\I \chi\pm 0) = \I y_0
+\text{p.v.} \int\limits_{\chi_c}^{1} \dfrac{\rho(\chi')\D \chi'}{\I
(\chi - \chi')} \pm \upi\rho(\chi), \ \chi_c<\chi<1.
\end{equation}Thus the jump of $\tilde z(\zeta)$ at branch cut is $-2\upi  \rho(\chi)$ for crossing branch cut at $\zeta=\I \chi$ in counterclockwise direction.

\subsection{Stokes wave as the sum of contribution from branch cuts in $w$  complex plane}

Consider a representation of Stokes wave by the density $\tilde\rho$
along branch cuts in complex plane $w$. Because of the
$2\upi$-periodicity in $u$ direction we write $z(w)$ as the integral
over periodically located branch cuts,
\begin{align} \label{fvdef}
 \tilde z(w) =z_1+\int\limits_{v_c}^{\infty}\sum\limits_{n=-\infty}^{\infty}  \left (\dfrac{1}{w +2\upi n- \I v'}-\dfrac{1}{b+2\upi n- \I v'}\right )\tilde \rho(v') \D v',
\end{align}
where $z_1$ is the complex constant, $v_c$ is related to $\chi_c$ by  \e{chicdef}, a summation over $n$ ensure the periodicity of $\tilde z(w)$ along $u$ and we replaced $\rho(\chi)$ by $\tilde \rho(v')$ to distinguish it from $\rho(\chi)$ in \e{branchcutdensity}.
Also we introduced the additional term $-\dfrac{1}{b +2\upi n- \I v'}$ which is intended to ensure the convergence of the integral. The constant $b$ can be chosen  at our convenience. A change of that constant results in the change of $z_1$.   %

The sum in \e{fvdef} is then  calculated using
of the identity%
\begin{align} 
\sum\limits^{\infty}_{n=-\infty}\frac{1}{n+a}=\upi\cot{\upi a}
\nonumber
\end{align}
giving
\begin{align} \label{fvdefsum}
 \tilde z(w) =z_1+\frac{1}{2} \int\limits_{v_c}^{\infty} \left (\cot{\left [\frac{w-\I v'}{2}\right ]}-\cot{\left [\frac{b-\I v'}{2}\right ]}  \right )\tilde\rho(v') \D v'.
\end{align}

Taking the limit $Im(w)\to -\infty$ we obtain from equations \e{fvdefsum} and  \e{branchcutdensity} that
\begin{align} \label{fvdefsumlimit}
 \tilde z(u-\I \infty) =z_1+\frac{1}{2} \int\limits_{v_c}^{\infty} \left ( i-\cot{\left [\frac{b-\I v'}{2}\right ]} \right )\tilde\rho(v') \D v'=iy_0+ \int\limits_{\chi_c}^{1} \dfrac{ \rho(\chi)\D \chi}{-\I -\I \chi}.
\end{align}
 We set
\begin{equation} \label{vpdef}
\chi=\tanh{\frac{ v'}{2}}
\end{equation}
and
\begin{align} \label{chivrelation}
\rho(\chi)=\tilde\rho(2\,\text{arctanh}\,{\chi}).
\end{align}


We also require that $z_1=\I  y_0$ then we find from the equations \e{fvdefsumlimit}, \e{vpdef} and \e{chivrelation}   that
\begin{equation} \label{cinf}
b=\upi.
\end{equation}

Using  the trigonometric identity
\begin{align} 
\cot{(a-b)}=\frac{1+\tan{a}\tan{b}}{\tan{a}-\tan{b}} \nonumber
\end{align}
one obtains from \eqref{fvdefsum}, \e{fvdefsumlimit}, \e{vpdef}, \e{chivrelation} and \e{cinf} that
\begin{align} \label{fvdefsum2}
\tilde z(w) = \I y_0+\frac{1}{2} \int\limits_{v_c}^{\infty}\left (
\dfrac{ \left[1+\I \tan{\frac{w}{2}}\tanh{\frac{ v'}{2}}\right ]
}{\tan{\frac{w}{2}}-\I \tanh{\frac{ v'}{2}}} -\I \tanh{\frac{ v'}{2}}
\right )\tilde\rho(v')\D v'\nonumber \\= \I y_0+\frac{1}{2}
\int\limits_{v_c}^{\infty} \left (\dfrac{ 1+\I \zeta\chi
}{\zeta-\I \chi}-\I \chi\right ) \tilde\rho(v')dv' =
\I y_0+\int\limits_{\chi_c}^{1} \dfrac{ \rho(\chi)\D
\chi}{\zeta-\I \chi},
\end{align}
i.e. we recovered the equation \e{branchcutdensity} from the equation \e{fvdef}.

\subsection{Expansion of $\rho(\chi)$ in powers of $\zeta-\I \chi_c$}
\label{sec:powerszetachic}

Assume that we have the branch cut $(\I \chi_c,\I )$ for $z(\zeta)$ in the complex plane of $\zeta$ and that the branch point at $\zeta=\I \chi_c$ is of square root type.
Then we expand $\rho(\chi)$ in the following series
\begin{align} \label{rhoseries1}
\rho(\chi)=\sum\limits_{n=0}^{\infty}b_{2n+1}(\chi-\chi_c)^{1/2+n}.
\end{align}
Note that that adding terms of integer powers of  $(\zeta-\I \chi_c)$ into the equation \e{rhoseries1} is not allowed because it would produce logarithmic singularity at $\zeta=\I \chi_c$ through the equation \e{branchcutdensity} which is incompatible with the Stokes wave as was shown in Refs. \citet{MalcolmGrantJFM1973LimitingStokes,TanveerProcRoySoc1991}.

Integrating over $\chi$ in \e{branchcutdensity}  using  \e{rhoseries1} gives
\begin{align} \label{fintsol1}
f(\zeta)&=b_1\left ( 2\I \sqrt{1-{\chi_c}}-2\I  \sqrt{\chi_c+\I  \zeta} \arctan\left[\frac{\sqrt{1-\chi_c}}{\sqrt{\chi_c+\I  \zeta}}\right]\right )\nonumber \\
&+b_3 \left( \frac{2}{3} \sqrt{1-\chi_c} (\I -4\I  \chi_c+3 \zeta)+2\I  (\chi_c+\I  \zeta)^{3/2}\arctan \left[\frac{\sqrt{1-\chi_c}}{\sqrt{\chi_c+\I  \zeta}}\right]\right )\nonumber \\
&+b_5\left ( \frac{2}{15} \sqrt{1-\chi_c} \left(3\I -11\I \chi_c+23\I \chi_c^2+5 \zeta-35 \chi_c \zeta-15 \I  \zeta^2\right)\right .\nonumber \\
&\qquad \qquad\left .-{2\I  (\chi_c+\I  \zeta)^{5/2}\arctan \left[\frac{\sqrt{1-\chi_c}}{\sqrt{\chi_c+\I  \zeta}}\right]}  \right )+b_7(\ldots)+\ldots
\end{align}

A series expansion of \e{fintsol1}
at $\zeta = \I \chi_c$ and comparison with the series   \e{halfser} result in the relations
\begin{align} \label{fintsol1ser}
b_{2j+1}=(-1)^{j+1}a_{2j+1}, \ j=0,1,2,\ldots
\end{align}
Note that   the expansion  \e{fintsol1} provides the relations for   $b_n$ with only  odd values of $n$. This is because  the series  \e{rhoseries1} is convergent only inside its   disk of convergence, $\chi-\chi_c<r$, where   $r$ is the radius of convergence. It will be shown in Part II that $r=2\chi_c$ for $\chi_c<1/3. $  The explicit expression for $\rho(\chi)$ is unknown for $\chi_c+r<\chi <1$ while  $\rho(\chi)$  still contributes to the terms $a_{2j}(\zeta-\I \chi_c)^{j},\ j=0,1,2,\ldots$ in the series  \e{halfser}.

Thus the expansion \e{rhoseries1} together with the relations \e{fintsol1ser} provides a convenient tool to work with $\rho(\chi)$ near to $\chi=\chi_c$.

\subsection{Absence of singularities in branch cut beyond the branch points $\zeta=\I \chi_c$ and $\zeta=\I.$}
\label{sec:absencebranchcut}

A priori one can not exclude existence of singularities inside the branch cut $\zeta\in [\I\chi_c,\I]$ beyond branch points $\zeta=\I\chi_c$ and $\zeta=\I$ at its ends. Existence of such singularities were conjectured in Refs. \citet{MalcolmGrantJFM1973LimitingStokes,SchwartzJFM1974}. To address that possibility we subtracted the expansion \e{fintsol1} from the numerical solution of $\tilde z(\zeta)$ for Stokes wave. We obtained both  $\tilde z(\zeta)$  and recovered $\rho(\chi)$ through AGH algorithm using variable precision arithmetics with $\sim 200$ digits  to achieve a high precision in that subtraction. Typically we used Stokes wave of the moderate nonlinearity with $\chi_c\sim 10^{-2}$  to operate with the moderate number of required Fourier harmonics. After that the numerical values  of $b_1,\, b_3, \, b_5,\ldots$ were recovered from fitting of $\rho(\chi)$ to the expansion \e{rhoseries1} near $\chi=\chi_c$. Typically we truncated the expansion \e{rhoseries1} to the first 3 terms $b_1,\, b_3, \, b_5$ which results in the truncated function $f(\zeta)_{truncated}$ in the expansion \e{fintsol1}. Also $\chi_c$ was obtained by the procedures described in Section \ref{sec:Findingchic}. Alternative way to recover  $b_1,\, b_3, \, b_5$  is through using the expansion \e{zkasymptotic2} was also used but generally gives lower precision.

Next step was to take $m$th derivative of   $\tilde z(\zeta)-f(\zeta)_{truncated}$ over  $\zeta$ numerically and obtain the Pad\'e approximation for the resulting expression    $\left [\tilde z(\zeta)-f(\zeta)_{truncated}\right ]^{(m)}$ resulting in new density $\tilde \rho(\chi)$. If any singularity would be present inside the branch cut then it would correspond to singularity in $\tilde \rho(\chi)$. However, we did not find any sign of such singularities at least for moderate order of derivative $m=1,2, 3$. It suggests that $\zeta=\I\chi_c$ and $\zeta=\I$ are the only singularities in complex $\zeta$ plain. This conclusion is in agreement with the results of both \citet{TanveerProcRoySoc1991} and Part II obtained by alternative methods.

\section{Conclusion}
\label{sec:Conclusion}

In this paper we found numerically the Stokes solutions of the primordial Euler
equations with free surface for large  range of wave heights,
including the approach to the limiting Stokes wave.  The limiting Stokes wave emerges as the singularity reaches
the fluid surface. We found from our high precision simulations (between 32 and more than 200 digits) the Pad\'e approximation of branch cut singularity of Stokes wave.  We provided the tables of  the Pad\'e approximants for a wide range of Stokes wave steepness. These tables allow to recover Stokes wave with the minimum accuracy $10^{-26}.$ We show that these  Pad\'e approximants quickly converge to  the jump at branch cut as the number of poles $N$ increases with the scaling law \e{errinf},\e{errscaling}.   We use the series expansion of the jump along branch cuts in half integer powers to recover the square-root singularity at the branch point. We found that there are no more singularities in the finite complex plane beyond one branch point per period. Following Part II is devoted to the analysis of the structure and location of branch points in infinite set of sheets of Riemann surface
beyond the physical sheet of Riemann surface considered here.


The authors would like to thank Prof.~S. Lau for the introduction to AGH method of Pad\'e
approximation and sharing his computer codes which were used at the initial stage of research. Also the authors thank
developers of FFTW~\citep{FFTW} and the whole GNU project~\citep{GNU}
for developing, and supporting this useful and free software. The
work of S.D. and A.K. was partially supported by the U.S. National
Science Foundation (grant no. OCE 1131791). The work of A.K. and
P.L. on the Pad\'e approximation was supported by Russian Science Foundation grant 14-22-00259.

\appendix
\section{Derivation of dynamical equations}
\label{DynDeriv} In this Appendix we  adapt the work
of~\citet{DKSZ1996}  to the case of the periodic boundary conditions deriving the basic  dynamical equations~\eqref{fullconformal1} and~\eqref{fullconformal2}
for
2D ideal hydrodynamics with free surface in conformal  variables. We  use similar  notations to \citet{DKSZ1996} and provide steps of the derivation skipped in \citet{DKSZ1996}.
\subsection{Hamiltonian after conformal map}
\begin{figure}
\centering
\includegraphics[width=3.5in]{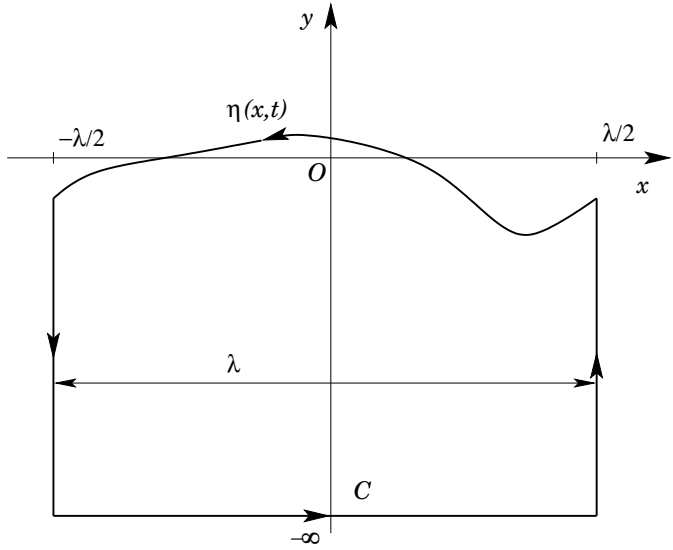}
\caption{\label{appendix_contour} A schematic of one period of a wave with a
counterclockwise contour of integration for application of Green's theorem.}
\end{figure}

It was shown by~\citet{Zakharov1968}, that the potential flow
of an ideal fluid  with free surface  is the canonical Hamiltonian system with canonical variables
$\eta$~\eqref{etadef} and $\psi$~\eqref{psidef}. Canonical Hamiltonian equations
\begin{equation}
\label{canonical_eqs} \frac{\p \eta}{\p t} = \frac{\delta
H}{\delta\psi},\;\; \frac{\p \psi}{\p t} = -\frac{\delta
H}{\delta\eta}.
\end{equation}
are equivalent to the boundary conditions~\eqref{kinematic1} and
\eqref{dynamic1}.
Here  $H$ is  the Hamiltonian which coincides with  the total energy (the sum of the kinetic energy $T$ and the potential energy $U$)  per spatial period of wave $\lambda$,
$$
H=T+U=\frac{1}{2}\int\limits_{-\lambda/2}^{\lambda/2}\D
x\int\limits_{-\infty}^{\eta}(\nabla \Phi)^2 \D y +
\frac{g}{2}\int\limits_{-\lambda/2}^{\lambda/2}\eta^2\D x,
$$
and without loss of generality the fluid density is set to one.   One has to express the kinetic energy
\begin{equation}\label{kineticdef}
T=\frac {1}{2}\int\limits_{-\lambda/2}^{\lambda/2}\D
x\int\limits_{-\infty}^{\eta}(\nabla \Phi)^2 \D y
\end{equation}
through the canonical variables  $\eta$ and $\psi$ which generally
requires to solve the Laplace equation \e{laplace} with the boundary
conditions \e{kinematic1}, \e{dynamic1}, \e{psidef} and
$\Phi(x,y,t)|_{y\to-\infty} = 0$ in the region $-\frac{\lambda}{2}
\le x < \frac{\lambda}{2},  \ -\infty < y \le \eta(x,t). $  That
region is schematically shown in Figure \ref{appendix_contour}.
 Using relations\begin{equation*}
\nabla \cdot(\Phi\nabla\Phi) = (\nabla\Phi)^2 +
\Phi\nabla^2\Phi, \quad (\nabla\Phi)^2 = \frac{\p}{\p
x}\left(\Phi\frac{\p\Phi}{\p x}\right) + \frac{\p}{\p
y}\left(\Phi\frac{\p\Phi}{\p y}\right),
\end{equation*}
which are valid for the harmonic function $\Phi$ \e{laplace} and applying Green's theorem to the equation \e{kineticdef} one  obtains that
\begin{equation}\label{Tsurfaceint}
2T=\int\limits_{-\lambda/2}^{\lambda/2}\int\limits_{-\infty}^{\eta}\left\{\frac{\p}{\p
x}\left(\Phi\frac{\p\Phi}{\p x}\right) + \frac{\p}{\p
y}\left(\Phi\frac{\p\Phi}{\p y}\right)\right\} \D x\D y =
\int\limits_{C} \left(-\Phi\frac{\p\Phi}{\p y}\D x +
\Phi\frac{\p\Phi}{\p x}\D y\right).
\end{equation}
Here $C$ is a positively (counterclockwise) oriented contour along the boundary of the
periodic domain occupied by fluid shown in Figure~\ref{appendix_contour}.
A sum of integrals along left and right hand sides (vertical segments) of the
contour  vanishes due to periodicity. Integral along
lower part of the contour (horizontal segment) is  zero due
to the boundary condition on potential $\Phi_{y=-\infty} = 0$. Notice
that in the case of finite depth fluid with a rigid flat bottom
$y=-h,$ the integral along similar lower segment $y=-h$  is
also zero because the boundary condition at the finite depth bottom  is
$\Phi_y|_{y=-h} = 0$ (zero vertical velocity at the  bottom) and $\D
y =0$. Then the equation \e{Tsurfaceint}
 is reduced to the following line integral\begin{equation}
\label{T_after_Green} 2T= \int\limits_{x=\lambda/2,\ y=\eta(x,t)}^{x=-\lambda/2}
\left(-\Phi\frac{\p\Phi}{\p y}\D x + \Phi\frac{\p\Phi}{\p x}\D
y\right).
\end{equation}

We use the time-dependent conformal transformation \e{ztconformal},\e{xL2} to relate partial derivatives in $x,y$ and $u,v$ as follows
\begin{eqnarray*}
&&\frac{\p \Phi}{\p u} = \frac{\p\Phi}{\p x}x_u + \frac{\p\Phi}{\p y} y_u,\\
&&\frac{\p \Phi}{\p v} = \frac{\p\Phi}{\p x}x_v + \frac{\p\Phi}{\p y} y_v ,\\
\end{eqnarray*}
which implies that
\begin{align}
\frac{\p \Phi}{\p x} = \frac{\Phi_u x_u - \Phi_v y_u}{x_u^2 + y_u^2},\label{Phi_x}\\
\frac{\p \Phi}{\p y} = \frac{\Phi_u y_u + \Phi_v x_u}{x_u^2 +
y_u^2},\label{Phi_y}
\end{align}
where we also used Cauchy-Riemann equations $x_u=y_v$ and $x_v=-y_u$ for
the conformal map $z(w)$.

Substituting~\eqref{Phi_x} and~\eqref{Phi_y}
into~\eqref{T_after_Green}, using relations $\D x = x_u \D u$ and
$\D y = y_u \D u$ on the line $w=u $ one obtains that\begin{equation}
\label{T_through_Phi_v} 2T = \int\limits_{\lambda/2}^{-\lambda/2}
\left .(-\Phi_v \Phi)\right |_{v=0} \D u = \int\limits_{-\lambda/2}^{\lambda/2} \left .\Phi_v
\Phi\right |_{v=0} \D u.
\end{equation}
Here we took into account the orientation of the contour and
conditions~\eqref{xL2} on conformal transformation.

Sokhotski–-Plemelj theorem \e{branchcutdensitySokhotski}  (see e.g. \cite{Gakhov1966,PolyaninManzhirov2008}) allows to   express a real part of the function
which is analytic in the lower (upper) half plane through the imaginary part
(and vice versa) at the real line $u=w$ using the Hilbert transformation~\eqref{Hilbertdef}.
For a conformal transformation $z(w,t)=x(w,t)+\I y(w,t)$ such relations are given
by~\eqref{xytransform}. A complex velocity potential
$\Pi(z,t) = \Phi + \I \Theta$ is the analytic function in the fluid domain $-\infty<y\le\eta(x,t)$, where $\Theta $ is the stream function. The conformal transformation $z=z(w,t)$ \e{ztconformal} ensures that $\Pi$  remains analytic function after transforming from $z$ to $w$ variable with the lower half
plane $\mathbb{C}^-$ being the domain of analyticity in $w$.  Similar to equation~\eqref{xytransform}, real and imaginary parts of $\Pi$ are related at the real line $u=w$ through the Hilbert transformation
as follows
\begin{equation} \label{PhiTheta1}
\Theta = \hat H \Phi, \quad \Phi =- \hat H \Theta,
\end{equation}
where we assumed the decaying boundary condition
 $\Phi |_{v=-\infty}=\Theta |_{v=-\infty} = 0.$ Here we abuse notation and use the same $\Pi$ and $\Phi$  both for independent variables $w$ and $z:$  $\tilde \Phi(w,t)\equiv \Phi(z,t)  $ and $\tilde \Pi (w,t)\equiv \Pi(z,t)$, i.e. we omit tilde.

Also the analyticity of $\Pi$ implies that the velocity
potential $\Phi$ is the harmonic function satisfying the Laplace equation \e{laplace} both in $x,y$ variables and similarly
$$ \nabla^2 \Phi (u,v,t)=0$$
in variables $u$ and $v.$  Using Cauchy-Riemann
equations \e{ThetauPhiv} and the relations \e{PhiTheta1} one obtains that
\begin{equation}
\label{Phi_v_Hilbert} \Phi_v = - \hat H
\Phi_u .
\end{equation}
Substituting~\eqref{Phi_v_Hilbert} into~\eqref{T_through_Phi_v} we
 express the kinetic energy in terms of canonical variable $\psi$ as follows
\begin{equation}
\label{T_through_Psi} 2T = \int\limits_{-\lambda/2}^{\lambda/2}
\left .\Phi_v \Phi \right|_{v=0} \D u = -\int\limits_{-\lambda/2}^{\lambda/2} \psi \hat H
\psi_u \D u,
\end{equation}
Here we used the definition~\eqref{psidef} which in $w$ plane turns into
$\psi(u,t) \equiv \Phi(u,v=0,t)$ as follows from the mapping of
the fluid surface into the real line $v=0$. Then the Hamiltonian in terms of variables on the
surface takes the following form
\begin{equation}\label{Huvariable}
H = -\frac{1}{2}\int\limits_{-\lambda/2}^{\lambda/2} \psi \hat H
\psi_u \D u +
\frac{g}{2}\int\limits_{-\lambda/2}^{\lambda/2}y^2x_u\D u.
\end{equation}
\subsection{Least action principle in conformal variables}
We use the constrained Lagrangian formulation to obtain the dynamical equations in conformal variables at fluid surface. A time dependence of the  map \e{ztconformal} implies  that we have to ensure the analyticity of that map through the appropriate constraint.   We discuss the Lagrangian dynamics first and add the corresponding constraint later in this Section. Equations~\eqref{canonical_eqs} realize
extremum of an action
\begin{equation}
\label{Action_general} S=\int\limits_{t_1}^{t_2} L \D t,
\end{equation}
with the Lagrangian
\begin{equation}
\label{Lagrangian_general} L=\int\limits_{-\lambda/2}^{\lambda/2}
\psi \frac{\p \eta}{\p t} \D x - H.
\end{equation}
The first term here has to be converted from the integral over $x$  into  $u$ variable. Consider
mapping $(x,t)\rightarrow (u,\tau)$, which is the change of parametrization of the surface under the conformal map.
Here  $\tau=t$.
Transformation $u=u(x,t)$ is the inverse to the conformal map
$x=x(u,\tau)$. The fluid surface $\eta(x,t)$ after transformation
corresponds to $y(u,\tau)$. We express $\p \eta/\p t$ by the chain rule as follows
\begin{equation}
\label{eta_t_1} \frac{\p\eta}{\p t} = \frac{\p
y}{\p\tau}\frac{\p\tau}{\p t} + \frac{\p y}{\p u}\frac{\p u}{\p t} =
\frac{\p y}{\p\tau} + \frac{\p y}{\p u}\frac{\p u}{\p t}.
\end{equation}
To find $\p u/\p t$ here we express full differentials of $x$ and $t$ through $u$ and
$\tau$ as follows
\begin{equation}
\left(
\begin{array}{c}
\displaystyle
\D x\\
\displaystyle \D t
\end{array}
\right) = \left(
\begin{array}{cc}
\displaystyle
x_u & x_\tau\\
\displaystyle t_u & t_\tau
\end{array}
\right)\left(
\begin{array}{c}
\displaystyle
\D u\\
\displaystyle \D \tau
\end{array}
\right) \equiv J \left(
\begin{array}{c}
\displaystyle
\D u\\
\displaystyle \D \tau
\end{array}
\right).
\end{equation}
Taking into account that $\tau=t$, one obtains the Jacobian matrix
\begin{equation}
\label{Jacobian_xt} J = \left(
\begin{array}{cc}
\displaystyle
x_u & x_\tau\\
\displaystyle 0 & 1
\end{array}
\right).
\end{equation}
Inverse procedure for full differentials of $u$ and $\tau$ through
$x$ and $t$ yields that
\begin{equation}
\label{u_tau_through_xt} \left(
\begin{array}{c}
\displaystyle
\D u\\
\displaystyle \D \tau
\end{array}
\right) = \left(
\begin{array}{cc}
\displaystyle
\p u/\p x & \p u/\p t\\
\displaystyle \p \tau/\p x & \p \tau/\p t
\end{array}
\right)\left(
\begin{array}{c}
\displaystyle
\D x\\
\displaystyle \D t
\end{array}
\right) = J^{-1} \left(
\begin{array}{c}
\displaystyle
\D x\\
\displaystyle \D t
\end{array}
\right).
\end{equation}
Comparing entries of the matrix in~\eqref{u_tau_through_xt} with
inverse of~\eqref{Jacobian_xt}, one gets
\begin{equation}
\label{u_by_t} \frac{\p u}{\p t} = -\frac{x_{\tau}}{x_u}.
\end{equation}
Substituting~\eqref{u_by_t} into~\eqref{eta_t_1} yields
\begin{equation}
\label{eta_t_2} \frac{\p\eta}{\p t} = y_{\tau} -
y_u\frac{x_\tau}{x_u}.
\end{equation}

We use the Lagrangian~\eqref{Lagrangian_general}  to substitute
it into the action~\eqref{Action_general}. Consider the first term in the action,
\begin{equation}
S = \int\int L \D t = \int\int \psi\frac{\p\eta}{\p t}\D x\D t +
\ldots
\end{equation}
and perform a change of variables in the integral as $\D x\D t = \det(J)\D u\D \tau
= x_u \D u\D \tau$. Together with the expression~\eqref{eta_t_2} it results in
\begin{equation}\label{firstterm}
\int\int \psi\frac{\p\eta}{\p t}\D x\D t = \int\int \psi
\left(y_{\tau} - y_u\frac{x_\tau}{x_u}\right)x_u \D u\D \tau =
\int\int \psi (y_{\tau}x_u - y_u x_\tau)\D u\D \tau.
\end{equation}
Using \e{Huvariable},  \eqref{Lagrangian_general}, \eqref{firstterm} and adding the analyticity constraint \e{xytransform} ensuring that $\tilde y -y_0=\hat H \tilde x$ and taking into account that $\tau=t$ as well we obtain a new constrained Lagrangian
\begin{align}
\label{Lagrangian_ut} L=\int\limits_{-\lambda/2}^{\lambda/2} \psi
(y_t x_u - y_u x_t)\D u +
\frac{1}{2}\int\limits_{-\lambda/2}^{\lambda/2} \psi \hat H \psi_u
\D u - \frac{g}{2}\int\limits_{-\lambda/2}^{\lambda/2}y^2x_u\D u
\\ +\int\limits_{-\lambda/2}^{\lambda/2}(y-\tilde y_0-\hat H \tilde x)f\D
u,\nonumber
\end{align}
where $f$ is the Lagrange multiplier for the analyticity
constraint.
\subsection{Variations of action}
We now obtain the dynamical equations from the Hamilton's least action principle. Vanishing of variational derivative  $\delta S/\delta \psi = 0$  of
the action~\eqref{Action_general} with the Lagrangian~\eqref{Lagrangian_ut}
over potential $\psi$ on the surface
yields the following expression\begin{equation}
\label{Var_by_psi} y_t x_u - y_u x_t + \hat H \psi_u = 0.
\end{equation}
This equation is nothing else but kinematic boundary
condition~\eqref{kinematic1} after the conformal map into $w$ plane.

Two conditions $\delta S/\delta x = 0$ and $\delta S/\delta y = 0$
result in  equations
\begin{align}
&y_u\psi_t - y_t\psi_u + g y y_u = \hat H f,\label{Var_by_x}\\
-&x_u\psi_t + x_t\psi_u - g yx_u = f,\label{Var_by_y}
\end{align}
which are turned into a single equation by excluding the Lagrange multiplier
$f$ giving
\begin{equation}
\label{Var_by_xy} y_u\psi_t - y_t\psi_u + \hat H (x_u\psi_t -
x_t\psi_u) + g[y y_u + \hat H(yx_u)] = 0.
\end{equation}
Equations \e{Var_by_psi} and \e{Var_by_xy} recover  the implicit dynamical equations~\eqref{fullconformal1} and~\eqref{fullconformal2}.

\subsection{Zeroth harmonic in implicit dynamical equations~\eqref{fullconformal1} and~\eqref{fullconformal2} and conservation of momentum}

Consider Fourier transformations of  the surface elevation $y(u,t)$  and the velocity potential on surface $\psi$
with respect to conformal coordinate $u$,
\begin{equation}
\label{Fourier_zeroth_harmonics}
\begin{array}{rl}
\displaystyle
&y(u,t) = y_0(t) + \sum\limits_{k\ne 0} y_k(t)\E^{\I ku},\\
\displaystyle
&\psi(u,t) = \psi_0(t) + \sum\limits_{k\ne 0} \psi_k(t)\E^{\I ku}.
\end{array}
\end{equation}
Here zeroth harmonics $y_0(t)$ and $\psi_0(t)$ are written separately and are given by
\begin{equation}
\label{zeroth_harmonics}
y_0(t)= \frac{1}{\lambda}\int\limits_{-\lambda/2}^{\lambda/2}y(u,t)\D u, \quad \psi_0(t)= \frac{1}{\lambda}\int\limits_{-\lambda/2}^{\lambda/2}\psi(u,t)\D u.
\end{equation}

One can rewrite equation \eqref{Var_by_xy} in the following form\begin{equation}
\label{Var_by_xy2}
y_u\psi_t - y_t\psi_u =- \hat H (x_u\psi_t -
x_t\psi_u + yx_u) - \frac{g}{2}\frac{\pD}{\pD u}y^2.
\end{equation}
A zeroth Fourier harmonic of the right hand side (r.h.s.) of equation \e{Var_by_xy2} vanishes because the term in parenthesis is multiplied by $\hat H$ which removes any zeroth harmonic and the remaining term is the partial derivative over $u.$ Respectively, the zeroth  harmonic of the left hand side (l.h.s.)
of equation  \eqref{Var_by_xy2}
must vanish.
Integrating that l.h.s.~ to obtain the zeroth harmonic, using equation  \e{Var_by_xy2} and integrating by parts over $u$ one obtains that
\begin{equation}
\label{zeroth_1}
\frac{1}{\lambda}\int\limits_{-\lambda/2}^{\lambda/2}(y_u\psi_t - y_t\psi_u)\D u =\frac{1}{\lambda}\int\limits_{-\lambda/2}^{\lambda/2}(y_u\psi_t + y_{ut}\psi)\D u =\frac{1}{\lambda}
\frac{\pD}{\pD t} \int\limits_{-\lambda/2}^{\lambda/2}\psi y_u\D u=0,
\end{equation}
 where we used a periodicity of $\psi$ and $y$ in $u.$ Thus $\int\limits_{-\lambda/2}^{\lambda/2}\psi y_u\D u$ is the integral of motion. To find a physical meaning of that integral we note that natural candidates for conserved
quantities are the components of the total momentum of  fluid along $x$ and $y$ directions. Taking into account that fluid density is one, we obtain  $x$ component of momentum $P_x$ as  an  integral of the horizontal velocity inside fluid, which gives
\begin{align}\label{Px}
P_x=\int\limits_{-\lambda/2}^{\lambda/2}\D
x\int\limits_{-\infty}^{\eta(x,t)}\Phi_x \D y = \int\limits_{C}\Phi\D y = \int\limits_{\lambda/2}^{-\lambda/2}\Phi\left .\frac{\pD y}{\pD x}\right|_{y=\eta(x,t)}\D x= -\int\limits_{-\lambda/2}^{\lambda/2}\psi y_u\D u.
\end{align}
Here we applied Green's theorem to positively oriented contour $C$ shown in Figure~\ref{appendix_contour}. Due to periodicity of functions and decaying
boundary condition  $\Phi(x,y,t)|_{y\to-\infty} = 0,$ only integral along the surface is nonzero.
Comparison of equations~\eqref{zeroth_1} and~\eqref{Px} shows that consistency of equation   \e{Var_by_xy2} is ensured by the conservation of the horizontal component  $P_x$ of the total momentum of the fluid.

Applying the Hilbert transformation $\hat H$ to equation~\eqref{Var_by_xy2} and using the identity \e{H2def0} one obtains that \begin{equation}
\label{Var_by_xy3}
x_u\psi_t - x_t\psi_u + yx_u  -q_0= \hat H \left(y_u\psi_t - y_t\psi_u + \frac{g}{2}\frac{\pD}{\pD u}y^2\right),
\end{equation}
where $q_0$ is the zeroth Fourier harmonic of $x_u\psi_t - x_t\psi_u + yx_u$.   To find $q_0$ we proceed similar to equations~\eqref{zeroth_1} and~\eqref{Px} to find that
\begin{equation}
\label{zeroth_2}
q_0=\frac{1}{\lambda}\int\limits_{-\lambda/2}^{\lambda/2}(x_u\psi_t - x_t\psi_u + yx_u  )\D u =\frac{1}{\lambda}
\frac{\pD}{\pD t} \int\limits_{-\lambda/2}^{\lambda/2}\psi x_u\D u+\frac{1}{\lambda}\int\limits_{-\lambda/2}^{\lambda/2} yx_u  \D u,
\end{equation}
 where
$\int\limits_{-\lambda/2}^{\lambda/2}\psi x_u\D u$ is the integral of motion corresponding to the conservation of the vertical component $P_y$ of the total momentum of  fluid,
\begin{align}\label{Py}
P_y=\int\limits_{-\lambda/2}^{\lambda/2}\D
x\int\limits_{-\infty}^{\eta(x,t)}\Phi_y \D y = \int\limits_{C}-\Phi\D x = \int\limits_{-\lambda/2}^{\lambda/2}\psi x_u\D u.
\end{align}
Then equations \e{zeroth_2} and \e{Py} imply that $q_0$ is the integral of motion given by
\begin{equation}
\label{q0a}
q_0=\frac{1}{\lambda}\int\limits_{-\lambda/2}^{\lambda/2} yx_u  \D
u=\frac{1}{\lambda}\int\limits^{\lambda/2}_{-\lambda/2}\eta(x,t)\D
x
\end{equation}
 and representing a conservation of the total mass of fluid. Also according to equation \e{yxucondition}, we set $q_0=0$ in this paper.

\section{Alpert-Greengard-Hagstrom (AGH) Algorithm and Stokes Wave}
\label{sec:AGHalgorithm} In this Appendix we  describe an
efficient algorithm for Pad\'e approximation of the function on a
discrete grid, following original work~\cite{AGH2000} and work
by~\cite{LauClassQuantumGrav2004} where more detailed explanation and further development of
the algorithm was presented.

Consider $2\upi$-periodic complex-valued function $f(u)=z(u)-u-\I y_0$ defined on a
 grid with nodes $u_j \in [-\upi,\upi]$. Values of the
function at the grid points are denoted as $f_j = f (u_j)$. We look
for an approximation of $f(u)$ in the form of a ratio of two
polynomials $P(u)$ and $Q(u),  $ i.e. the Pad\'e approximation. We
briefly describe AGH algorithm in a general way with additional
comments for our particular case. As it was mentioned in
Section~\ref{sec:PadeApproximationStokesWave}, we use the second
conformal map $\zeta = \tan(u/2)$. The introduction of auxiliary
variable $\zeta$ allows to consider the real line
$\zeta\in\mathbb{R}$ as opposed to considering a finite interval
$u\in[-\upi,\upi]$, while the infinity along the imaginary axis is
mapped into imaginary unit $\I$ and  $2\upi$-periodicity in $u$
direction is ensured. Without loss of generality we assume that
$f(\pm\upi)=0$. In  this paper, we take  $f(u)=z(u)-u-\I y_0=\tilde
z(u)-\I y_0 $  (see equations \e{branchcutdensity} and
\e{approx_cut} for comparison). Condition $f(\pm\upi)=0$ allows  to
consider $P$ and $Q$ such that the degree of polynomials are
$\deg{Q} = 1+\deg{P} = N$, where the integer  $N$ is allowed to
vary. We are looking for the convergence of the rational
approximation to $f$, \begin{align*} \dfrac{P(\zeta)}{Q(\zeta)} \to
f(\zeta),
\end{align*}
in a sense  of solving a minimization problem
\begin{equation}
\label{AP1:nonlinear_gen}
 \min\limits_{P,Q} \int\limits_{-\upi}^{+\upi} \left| \dfrac{P(u)}{Q(u)}
- f(u) \right|^2 {\D u}.
\end{equation}
That minimization problem  is challenging  because $Q$ in the denominator  makes   \e{AP1:nonlinear_gen}    nonlinear problem.
In  the transformed variable $\zeta$ the problem  \e{AP1:nonlinear_gen}  remains nonlinear and  is reduced to
\begin{equation}
\label{AP1:nonlinear}
 \min\limits_{P,Q} \int\limits_{-\infty}^{+\infty} \left| \dfrac{P(\zeta)}{Q(\zeta)}
- f(\zeta) \right|^2 \dfrac{\D \,\zeta}{\zeta^2+1}.
\end{equation}
In AGH algorithm, the complexity of nonlinearity is bypassed by solving instead
of~\eqref{AP1:nonlinear_gen}, a sequence of  linear least-square problems
\begin{equation}
\label{AP1:linearized}
 \min\limits_{P^{(i+1)},Q^{(i+1)}} \int\limits_{-\infty}^{+\infty} \left| \frac{P^{(i+1)}(u)}{Q^{(i)}(u)}
- \frac{Q^{(i+1)}(u)}{Q^{(i)}(u)} f(u) \right|^2 \D u, \ i=1,2,\ldots
\end{equation}
We define an inner product
\begin{equation}
\langle f, g\rangle_i = \int\limits_{-\infty}^{+\infty} f(u) \bar
g(u) w_i(u) \D u,
\end{equation}
 with a weight function $w_i (u)= \frac{1}{|Q^{(i)}(u)|^2}$ (for  $\zeta$-plane the formula for the weight is modified to be $w_i(\zeta)= 1/(|Q^{(i)}(\zeta)|^2 (\zeta^2+1))$). Then the previous least squares problem can be rewritten as follows\begin{equation}
\label{AP1:linearized_inner}
 \min\limits_{P^{(i+1)},Q^{(i+1)}} || - P^{(i+1)}(u)
+ Q^{(i+1)}(u)f(u) ||_i,
\end{equation}
where  $$
|| g(u) ||^2_i = \langle g, g \rangle_i
$$
is the norm.

As it was shown in~\cite{AGH2000}, the solution of the least squares
problem~\eqref{AP1:linearized_inner} is equivalent to the solution of
$$
\langle - P^{i+1} + Q^{i+1} f(u), h_n(u)\rangle_i = 0,
$$
for $n=1,\ldots,2N$, with $h_n(u)$ defined as follows
\begin{equation}
\begin{cases}
u^{n/2-1}, & \mbox{for } n=2,4,6,\ldots,2N,\\
u^{(n-1)/2} f(u), & \mbox{for } n=1,3,5,\ldots,2N-1,\\
\end{cases}
\end{equation}
which are nothing else but
$$
f(u), 1, u f(u), u, u^2 f(u),\ldots, u^{N-1} f(u), u^{N-1}, u^N
f(u).
$$
This claim can be proven by variation of $m$th coefficient of
$P(u)$ (for even $n=2m+2$) and $Q(u)$ (for odd $n=2m+1$). We put
coefficient at the leading power of $Q(u)$ to be equal to one. Thus
we need to find $2N$ coefficients for two polynomials.

We orthogonalize $2N+1$ functions $h_n(u)$ using Gramm-Schmidt
orthogonalization procedure,\begin{equation}
g_n(u)=
\begin{cases}
f(u), & \mbox{for } n=1,\\
1-c_{21}f(u), & \mbox{for } n=2,\\
u g_{n-2}(u) - \sum\limits_{j=1}^{\min\{4,n-1\}}c_{nj}g_{n-j}(u), &
\mbox{for } n=3,\ldots,2N+1,
\end{cases}
\end{equation}
where real constants $c_{nj}$ are given by
\begin{equation}
c_{nj}=
\begin{cases}
\frac{\langle 1, f(u)\rangle_i}{\langle f(u), f(u)\rangle_i}, & \mbox{for } n=2\,\mbox{and } j=1,\\
\frac{\langle (u g_{n-2}, g_{n-j}\rangle_i}{\langle g_{n-j},
g_{n-j}\rangle_i}, & \mbox{for } n=3,\ldots,2N+1\,\mbox{and }
j=1,\ldots,\min\{4,n-1\}.
\end{cases}
\end{equation}
Then we obtain that
\begin{equation}
g_{2N+1} = -P^{(i+1)} + f(u)Q^{(i+1)},
\end{equation}
so $P^{(i+1)}$ and $Q^{(i+1)}$ are computed from recurrence
coefficients $c_{nj}$ by splitting into even and odd-numbered
parts.

For our purposes of finding the jump at branch cut it is convenient to represent a ratio of $P(u)$ and
$Q(u)$ as a sum of simple poles,\begin{equation}
\frac{P(u)}{Q(u)} = \sum\limits_{n=1}^{N}\frac{\gamma_n}{u -
\chi_n}.
\end{equation}
In order to do that we compute  zeros  $\chi_n,\ n=1,2,\ldots, N$ of $Q(u)$ using
Newton's iterations. At each step one zero $\chi_n$ is found by Newton's iterations. After that we remove that zero from $Q(u)$ by division on $(u-\chi_n)$ and proceed to the next step for the modified $Q$ etc. After
that procedure coefficients $\gamma_n$ are given by the following expression
\begin{equation}
\gamma_n = \frac{P(\chi_n)}{Q'(\chi_n)}.
\end{equation}
Derivative $Q'(u)$ are obtained from previous recurrence relation
for $g_n(u)$ by differentiation.

\section{Tables of Stokes Waves}
\label{sec:TablesStokesWaves}

Using the Pad\'e approximation, introduced in
Section~\ref{sec:PadeApproximationStokesWave}, one can approximate
Stokes wave for each value of the scaled height $H/\lambda$ as a sum
of poles
\begin{equation}
\label{Pade_formula} z(w) \simeq z_{pade}(u)\equiv w + \I y_0 +
\sum\limits_{n = 1}^{N} \dfrac{\gamma_n}{\tan(w/2) - \I\chi_n}.
\end{equation}
Here $N$ is the number of poles in the Pad\'e approximation. Using AGH algorithm
(see Appendix~\ref{sec:AGHalgorithm}) we found that all
 poles for all values of  $H/\lambda$  are located on the imaginary axis.

We provide Tables \ref{tab:c1.005}-\ref{tab:c1.0922851405} for four
particular cases of Stokes waves with wave heights ranging from $H/L
\simeq 0.031791$ to $H/L \simeq 0.141058$.
 Complete library of computed
waves can be accessed through the electronic attachments as well as
through the web link~\cite{PadePolesList}.
These data of Pad\'e approximation allow to recover the Stokes wave
with the relative accuracy of at least $10^{-26}$ (for the vast majority
of cases the actual accuracy is higher by  several orders of magnitude). First and second
columns of both Tables and electronic files represent values of
$\chi_n$ and $\gamma_n$, respectively.  Additionally a  third column in electronic files provides the values of $\rho_{n,N},\ n=1,2,\ldots,N$ calculated from data of the first two columns using equations \eqref{rhoka}, \e{rho1an} and \e{rhoNN}.

We used three quantities to characterize the accuracy of our numerical Stokes wave solution and its Pad\'e approximation. First quantity is the residue
$$
R(y) \equiv N^{-1/2} \left({\sum\limits_{j=1}^M |\hat L_0 y(u_j)|^2}\right )^{1/2}
$$
of equation~\eqref{stokes_wave2}. $R(y)$ characterizes convergence of our iteration algorithm described in Section \ref{sec:NewtonCG}  to the Stokes wave. Here $M=2k_{max}$ is the number of grid points $u_j$ used in the discretization of $z(u). $    Second quantity is the relative error of  Pad\'e approximation
$$
err_{pade} = \left({\frac{\sum\limits_{ j=1} ^M|z(u_j) -
z_{pade}(u_j)|^2}{\sum\limits_{ j=1}^M|z_{}(u_j)|^2}}\right )^{1/2}
$$
of our numerical solution $z(u_j$).
Third quantity is the amplitude of the highest Fourier harmonics $|\hat z_{k_{max}}|$ used in FFT.

We balanced these three quantities in our simulation to achieve the
most efficient and reliable approximants of Stokes waves. Typically
we chose $k_{max}$ large enough such that $|\hat
z_{k_{max}}|<M^{-1/2}10^{-26}$ to ensure that our discretization
error is below $10^{-26}$. Here  the factor $M^{-1/2}$ characterize
the accumulation of round-off error in FFTs. A convergence of
numerical iterations down to $R(y)\simeq 10^{-28}$ was found to be
sufficient to achieve the desired accuracy of solution in $10^{-26}$.
After that we used AGH algorithm with $N$ large enough to make sure
that $err_{pade}$ is below $10^{-26}$ by several orders of
magnitude.

The  second and third rows in electronic {\tt .dat}-files provide
the additional information extracted from simulations which include
the number of points of the numerical grid $M=2k_{max}$, the
residual $R(y),$ the Stokes wave height $y_0$ at $x=\pm \upi$, the
amplitude of the highest Fourier harmonics $|\hat z_{k_{max}}|$, the
Pad\'e error $err_{pade}$, the scaled Stokes wave height $H/\lambda$
and the Stokes wave velocity $c.$ Values of $H/\lambda$ are also
encoded in the names of {\tt .dat}-files. Also the file {\tt
summary.txt} provides a summary of the results from all {\tt
.dat}-files.

\begin{table}
\begin{center}
\def~{\hphantom{0}}
\begin{tabular}{lccc}
$k$  & $\chi_k$ & $\gamma_k$ \\
1   & 9.96041092606335083862992746108661e-01 & 7.86955267798815779896940975384730e-03\\
2   & 9.78972925087544517288851755005498e-01 & 1.58938208649989549970220156007558e-02\\
3   & 9.49569603918982534434611327659588e-01 & 2.09270477914666067462762444957813e-02\\
4   & 9.10406118678767801011884022998371e-01 & 2.30772855309232774921927032978593e-02\\
5   & 8.64694768844023775849318292632706e-01 & 2.27821823395535616299282467454014e-02\\
6   & 8.15784392644967788264370902239031e-01 & 2.06781569334250898322582073617042e-02\\
7   & 7.66774518804464747111211133286936e-01 & 1.74505018291868390667201976549862e-02\\
8   & 7.20283901206785220595281417979068e-01 & 1.37191518620555270387052855587868e-02\\
9   & 6.78365420413127130751057141705342e-01 & 9.97903489393458824811047018470610e-03\\
10   & 6.42527484790841967661585477231496e-01 & 6.58739944600110438403394267024205e-03\\
11   & 6.13814667765069562012985702014292e-01 & 3.78022620936003042036390618792491e-03\\
12   & 5.92908315774231571020102078418028e-01 & 1.70068438916655758680561664969251e-03\\
13   & 5.80220882639295372104402613045357e-01 & 4.27939536191998511898005177240895e-04\\
\end{tabular}
\caption{Data for Pad\'e approximation of the wave with velocity $c = 1.005$, the steepness $H/\lambda=0.031791185830078550217424174610939$, and $y_0 = -0.094819818875344225940453182945545$. Parameters of simulations and Pad\'e approximation are $M = 16384$, $R(y)\simeq 3.64\times 10^{-33}$, $ err_{pade} \simeq 4.65\times 10^{-31}$, and the smallest Fourier harmonic had value $|\hat z_{k_{max}}| \simeq 1.00\times 10^{-39}$.}
\label{tab:c1.005}
\end{center}
\end{table}

\begin{table}
\begin{center}
\def~{\hphantom{0}}
\begin{tabular}{lccc}
$k$  & $\chi_k$ & $\gamma_k$ \\
1   & 9.95104877443162988543285604210300e-01 & 9.86344259137750131929660816853428e-03\\
2   & 9.74036453796113160099814623502153e-01 & 2.04507668432927246329100238980833e-02\\
3   & 9.37570348097817646693771937993553e-01 & 2.79392155141519735983551710052556e-02\\
4   & 8.88487837568099082583936075213862e-01 & 3.23466950514503536891484865775305e-02\\
5   & 8.30226745019764721513358604279726e-01 & 3.40147781594254640293216855633111e-02\\
6   & 7.66333992991599455804305516283483e-01 & 3.35031653839781443242921103012960e-02\\
7   & 7.00044654389407424698796692053902e-01 & 3.14489535896823273634285653396772e-02\\
8   & 6.34030062620564487061621925382604e-01 & 2.84500818501058994363745193170182e-02\\
9   & 5.70306175038391450182479612388228e-01 & 2.49972283728050063777939045774063e-02\\
10   & 5.10259067527902434999804572718914e-01 & 2.14511658748252720344184761694938e-02\\
11   & 4.54736803912594713785686635920502e-01 & 1.80506679368664453110659093724408e-02\\
12   & 4.04166063099136862080187337641328e-01 & 1.49353242488736140402002871952586e-02\\
13   & 3.58666569676145947416712220270910e-01 & 1.21719831322217078183473222652247e-02\\
14   & 3.18149576080072883402812844826148e-01 & 9.77859794087821305028150640251530e-03\\
15   & 2.82395821449537219671027831006306e-01 & 7.74304960914297497673324174361282e-03\\
16   & 2.51113624911206124013006119628666e-01 & 6.03673166180751660604462350895498e-03\\
17   & 2.23980136883660703565243361779229e-01 & 4.62368871094252860720889700284368e-03\\
18   & 2.00669410169964201214033391681290e-01 & 3.46637996984113299044046402602488e-03\\
19   & 1.80870715726851513117763940777923e-01 & 2.52906751261377992038440816275726e-03\\
20   & 1.64299946133899344356900387886524e-01 & 1.77962987225519971005973434397249e-03\\
21   & 1.50706307186061197360270221491920e-01 & 1.19038860012342014795358493343618e-03\\
22   & 1.39875923302152112463866847539994e-01 & 7.38354660230910637368329193600582e-04\\
23   & 1.31633517886184493678824771993353e-01 & 4.05164363997062298811235878087573e-04\\
24   & 1.25842975665215746611096122396649e-01 & 1.76877312168739070195683856144061e-04\\
25   & 1.22407333541749966033626434463374e-01 & 4.37430353029634816661994533800963e-05\\
\end{tabular}
\caption{Data for Pad\'e approximation of the wave with velocity $c = 1.051$, the steepness $H/\lambda=0.10042675172528485854673515635249$, and $y_0 = -0.25732914098527682158156915646871$. Parameters of simulations and Pad\'e approximation are $M = 16384$, $R(y)\simeq 5.19\times 10^{-32}$, $ err_{pade} \simeq 1.69\times 10^{-31}$, and the smallest Fourier harmonic had value $|\hat z_{k_{max}}| \simeq 1.00\times 10^{-37}$.}
\label{tab:c1.051}
\end{center}
\end{table}

\begin{table}
\begin{center}
\def~{\hphantom{0}}
\begin{tabular}{lccc}
$k$  & $\chi_k$ & $\gamma_k$ \\
1   & 9.95433825932608550132384034857213e-01 & 9.06513968659594263994154983859022e-03\\
2   & 9.75720732729872361661639075445556e-01 & 1.88104172479867625607988377665013e-02\\
3   & 9.41458580699036796998398974950063e-01 & 2.58024268240062770737774505111002e-02\\
4   & 8.95072079198109934865810125596955e-01 & 3.00760315302764871790369706833779e-02\\
5   & 8.39603368860036006278810820983949e-01 & 3.19167851791100543244698807489031e-02\\
6   & 7.78246707846125864530269738095348e-01 & 3.17900776761535529310595879964877e-02\\
7   & 7.13977463595480003911919153327480e-01 & 3.02335467307095385878230877915997e-02\\
8   & 6.49314375127269358025790723963870e-01 & 2.77625384158746106085149228773401e-02\\
9   & 5.86215465928989576951120366245581e-01 & 2.48107842884599707048520815911264e-02\\
10   & 5.26077937447751872984275728101522e-01 & 2.17068351144352237760178371853073e-02\\
11   & 4.69802407403950485965504029996914e-01 & 1.86762941525117699745376670289500e-02\\
12   & 4.17886269376110040045868819592273e-01 & 1.58579969144004349124961258254260e-02\\
13   & 3.70521444182340763884420027666827e-01 & 1.33248510736740059585812992162263e-02\\
14   & 3.27682439755960213510467363606042e-01 & 1.11037283880977232781708686935009e-02\\
15   & 2.89198722549172239607721584343698e-01 & 9.19185285105063318623096617861130e-03\\
16   & 2.54810466354666195816665370686997e-01 & 7.56907825149316224628365974410675e-03\\
17   & 2.24209367334721589574533037466343e-01 & 6.20644453419544059242384682030945e-03\\
18   & 1.97067225067065397541246930106441e-01 & 5.07177162315234989673990118919696e-03\\
19   & 1.73055091362268451120364081334665e-01 & 4.13307962854491967554778057762908e-03\\
20   & 1.51855462026532436104149361379210e-01 & 3.36050786167674783997799826083550e-03\\
21   & 1.33169515540203788156642166906591e-01 & 2.72724894316322072325277221844458e-03\\
22   & 1.16720933274701291000887725539493e-01 & 2.20986855953690886419588367706623e-03\\
23   & 1.02257431405919038178275031736290e-01 & 1.78826390264321999398279771128206e-03\\
24   & 8.95508127283858594116407019601820e-02 & 1.44542657328550726857080252749622e-03\\
25   & 7.83961028319930614767173088020279e-02 & 1.16711442195804019558665126539427e-03\\
26   & 6.86101566854466616210976539289158e-02 & 9.41495432894175937842874927421205e-04\\
27   & 6.00299941184905270702652324982135e-02 & 7.58799737152752549854592147665995e-04\\
28   & 5.25110331386998700818247974509258e-02 & 6.10998699362372783997691390696487e-04\\
29   & 4.59253281231247667007365599867943e-02 & 4.91519483298208817983041233383980e-04\\
30   & 4.01598777968588267195919995947710e-02 & 3.94997269772740759120747730149340e-04\\
31   & 3.51150396988382900972228874525109e-02 & 3.17063815302569437242780768838413e-04\\
32   & 3.07030693071633025856068326424687e-02 & 2.54169263655189066241987434556100e-04\\
33   & 2.68467901323811952321185618074578e-02 & 2.03433380026332773186591672337583e-04\\
34   & 2.34783937316708681288640025635146e-02 & 1.62522237707382077096302354822532e-04\\
35   & 2.05383642178678839820064784655908e-02 & 1.29546576264629788075276452826328e-04\\
36   & 1.79745193852614117777103084010265e-02 & 1.02978399018687148172885868196623e-04\\
37   & 1.57411593704806090331919997928406e-02 & 8.15827862763857084235776103594481e-05\\
38   & 1.37983133839445750806859736203507e-02 & 6.43623147351034758281773572995289e-05\\
39   & 1.21110752013028358188007191935072e-02 & 5.05118646282732166085507134151195e-05\\
40   & 1.06490185910779922966530559824050e-02 & 3.93819507325944359184298112513636e-05\\
41   & 9.38568452652012956402075369290278e-03 & 3.04490256349179913250724576956079e-05\\
42   & 8.29813278439420925351354881096343e-03 & 2.32914730544885187107604807304053e-05\\
43   & 7.36655130339489438745177566237321e-03 & 1.75702379530898436181417357806391e-05\\
44   & 6.57391741813790189283510916305043e-03 & 1.30132325361793585312305562950935e-05\\
45   & 5.90570578080460414802523913159400e-03 & 9.40281755981180973510534726928533e-06\\
46   & 5.34963842759485046283496838777743e-03 & 6.56579130822932954380944114441017e-06\\
47   & 4.89547304507788970804144262729396e-03 & 4.36542870899759471322771653831421e-06\\
48   & 4.53482604688247134485690783763108e-03 & 2.69520431445801848314129686051970e-06\\
49   & 4.26102758786433892841273275905455e-03 & 1.47390871215592819437170943388163e-06\\
50   & 4.06900612679011263031777402788905e-03 & 6.41931271126488105657466176599801e-07\\
51   & 3.95520060861743412410650474994064e-03 & 1.58535598642490907942230532511450e-07\\
\end{tabular}
\caption{Data for Pad\'e approximation of the wave with velocity $c = 1.0929$, the steepness $H/\lambda=0.13825830866311310404416736817381$, and $y_0 = -0.2915339172431288292999965032009$. Parameters of simulations and Pad\'e approximation are $M = 65536$, $R(y)\simeq 2.59\times 10^{-31}$, $ err_{pade} \simeq 1.01\times 10^{-32}$, and the smallest Fourier harmonic had value $|\hat z_{k_{max}}| \simeq 5.00\times 10^{-38}$.}
\label{tab:c1.0929}
\end{center}
\end{table}

\begin{table}
\begin{center}
\def~{\hphantom{0}}
\begin{tabular}{lccc}
$k$  & $\chi_k$ & $\gamma_k$ \\
1   & 9.93398643583003025153415435504531e-01 & 1.28741415762741679346145664996829e-02\\
2   & 9.65060482058669453480870980252106e-01 & 2.60267721567181859054226183119919e-02\\
3   & 9.16714789789161849202896809849705e-01 & 3.43697679627825352269776411757291e-02\\
4   & 8.53133231672953078297597841861823e-01 & 3.81940153445550033591099941584920e-02\\
5   & 7.79861194757267505357891461462460e-01 & 3.83579555843352602595202964927154e-02\\
6   & 7.02176870736814955937036169062899e-01 & 3.59774931820106663192752740784546e-02\\
7   & 6.24431910013545641111933110534714e-01 & 3.21282856711997437340001627375168e-02\\
8   & 5.49795332847499273525022520666518e-01 & 2.76691730121067711930378384900232e-02\\
9   & 4.80296232546974823630387324524545e-01 & 2.31906560517515347893552651709825e-02\\
10   & 4.17024645857800791824977075950470e-01 & 1.90429324287558759377242844714520e-02\\
11   & 3.60378305224974615221811203424356e-01 & 1.53959152585563261217878597243142e-02\\
12   & 3.10289842646985418345605736936048e-01 & 1.23006351687210032636876990145911e-02\\
13   & 2.66407648286693861932841258994721e-01 & 9.73852232172353639036354649203610e-03\\
14   & 2.28226886286517332438415939592608e-01 & 7.65583101348162009153247587551195e-03\\
15   & 1.95177843545134733789119725180096e-01 & 5.98535170044127311114464478095979e-03\\
16   & 1.66681805023593130823224346797399e-01 & 4.65887488461708640776107914945203e-03\\
17   & 1.42183993584453030868317439497003e-01 & 3.61357577121517253016005783742974e-03\\
18   & 1.21171172484979583446779037786714e-01 & 2.79470260489814969664959294926652e-03\\
19   & 1.03179452511723992079582272565214e-01 & 2.15617836676118577227243251241331e-03\\
20   & 8.77961228595568606401648108610398e-02 & 1.66012679310648786275121618740345e-03\\
21   & 7.46580315447597465030574978595406e-02 & 1.27592107645668605269203574613739e-03\\
22   & 6.34481306443940719507463127666211e-02 & 9.79089087450639485345366472519338e-04\\
23   & 5.38911872290903942028496746096495e-02 & 7.50248095624531518737269429111365e-04\\
24   & 4.57492587887819300483008251206986e-02 & 5.74148509163471182463355721181721e-04\\
25   & 3.88172752742240105088258746631741e-02 & 4.38854383779301056123215099079188e-04\\
26   & 3.29189096590597195144947958198357e-02 & 3.35061475940689863018834966768472e-04\\
27   & 2.79028212558501660025568098114343e-02 & 2.55540871176311150876065830758264e-04\\
28   & 2.36392981742008042750164949870111e-02 & 1.94691329540109355414952726522057e-04\\
29   & 2.00172924085211268884110103807234e-02 & 1.48182722053010736525844218770677e-04\\
30   & 1.69418235041002160106748854919734e-02 & 1.12674155188771601572543993641392e-04\\
31   & 1.43317184931477833930656499987995e-02 & 8.55924207962470484084537963575900e-05\\
32   & 1.21176530092033068365142201489852e-02 & 6.49586463790799447674548874436684e-05\\
33   & 1.02404588173178159766150762312400e-02 & 4.92531498561398293539370127249816e-05\\
34   & 8.64966499042977177325090332415729e-03 & 3.73103914016284265439688122490895e-05\\
35   & 7.30224274879133050003140619052446e-03 & 2.82375225733516039477667066021446e-05\\
36   & 6.16152705009717845830463720403643e-03 & 2.13513661747602139469739101816769e-05\\
37   & 5.19629108350495067218636168760723e-03 & 1.61297466879220448079982238240246e-05\\
38   & 4.37995272297094390520221610919454e-03 & 1.21739653477311336991048098474804e-05\\
39   & 3.68989465744363535075551189083628e-03 & 9.17991094830884893227377698101798e-06\\
40   & 3.10688231013106004394611827804953e-03 & 6.91584935818872179036239759592572e-06\\
41   & 2.61456578557885722141496603598660e-03 & 5.20536938490202499152488268561648e-06\\
42   & 2.19905395393225001547396749458147e-03 & 3.91430351823738560080383189178292e-06\\
43   & 1.84855041619759608964615915855160e-03 & 2.94070847979427725012698794615673e-06\\
44   & 1.55304251639257745230445095694940e-03 & 2.20719808861136536998364371692094e-06\\
45   & 1.30403580024761683344961061119599e-03 & 1.65508232204597074734186170013816e-06\\
46   & 1.09432738764763509113618247317394e-03 & 1.23989163441506187230543611758337e-06\\
47   & 9.17812647645417749658202151298786e-04 & 9.27962520159008244607471765496508e-07\\
48   & 7.69320359699922614599799922262156e-04 & 6.93835226000040035242065190155287e-07\\
49   & 6.44472229750674596538655661872796e-04 & 5.18272340903010891482796683060429e-07\\
50   & 5.39563219268408452082296554789566e-04 & 3.86751562364387865498882802928547e-07\\
51   & 4.51459652187210048535645581625337e-04 & 2.88320243535854856966320396641389e-07\\
52   & 3.77512500082907183102947952504772e-04 & 2.14725698524118689197774299181126e-07\\
53   & 3.15483620098192985576130479232556e-04 & 1.59755495420848083506690922212921e-07\\
54   & 2.63483041310020855659728424130368e-04 & 1.18737501808356300532888071897965e-07\\
55   & 2.19915670752623096428691515132722e-04 & 8.81613507227033650558982811933814e-08\\
\end{tabular}
\end{center}
\end{table}
\newpage\begin{table}
\begin{center}
\def~{\hphantom{0}}
\begin{tabular}{lccc}
$k$  & $\chi_k$ & $\gamma_k$ \\
56   & 1.83436026617218827592565025305802e-04 & 6.53921058513165146118265854353209e-08\\
57   & 1.52909808789403664334781521734162e-04 & 4.84538716133603951388247789431925e-08\\
58   & 1.27381290492618707318705788922866e-04 & 3.58664158964311161459118236867907e-08\\
59   & 1.06045663430404223545931459157107e-04 & 2.65219345830887370991540659742850e-08\\
60   & 8.82255960623097328938047290772380e-05 & 1.95921833026360368361816879242692e-08\\
61   & 7.33513735582244917463595672161340e-05 & 1.44585603393432845280965460758372e-08\\
62   & 6.09440811020089107359438382849930e-05 & 1.06595193757093831390156057934750e-08\\
63   & 5.06013718010847795103344003355303e-05 & 7.85105510061063345075831379965022e-09\\
64   & 4.19854284842886757235756347132190e-05 & 5.77704094551584320668305953789402e-09\\
65   & 3.48127867887149603598695647961378e-05 & 4.24698455656734771613508619208412e-09\\
66   & 2.88457365344826502539670719441546e-05 & 3.11936272695846692633031613888051e-09\\
67   & 2.38850607015635551072991988129380e-05 & 2.28914896008025877293425583671707e-09\\
68   & 1.97639074377385155071521574022480e-05 & 1.67848842336096800302384900691726e-09\\
69   & 1.63426213283602794258356970211462e-05 & 1.22973329115351088076908027814457e-09\\
70   & 1.35043863909173156500338815494453e-05 & 9.00246494394826687766241103823572e-10\\
71   & 1.11515555862156824939722316951789e-05 & 6.58529018137743573464498106258884e-10\\
72   & 9.20256064170640198246888785687659e-06 & 4.81336779425064219659348751354589e-10\\
73   & 7.58931213612359462965716406185546e-06 & 3.51536622960117080648972753857693e-10\\
74   & 6.25501349165233025416447266351498e-06 & 2.56513789011984140244514391180857e-10\\
75   & 5.15232409179580442223607363260130e-06 & 1.86990411531843046778670738725091e-10\\
76   & 4.24181649775016522927709678588896e-06 & 1.36150027070583877598207373400005e-10\\
77   & 3.49068106812952109834665722037669e-06 & 9.89896125536513948820894793870582e-11\\
78   & 2.87163870904281978123381120499882e-06 & 7.18405497951763524448150948318652e-11\\
79   & 2.36202927278644614134938592885535e-06 & 5.20148665829541462756399079612847e-11\\
80   & 1.94304885285494843909466767533574e-06 & 3.75443719949043320491081089015406e-11\\
81   & 1.59911321065017380641753434790850e-06 & 2.69886771259819847047462342227678e-11\\
82   & 1.31732715240169671433477796333342e-06 & 1.92941913934894250545144244154976e-11\\
83   & 1.08704211469651017023210815498690e-06 & 1.36906826573455708305877340415078e-11\\
84   & 8.99487103028405911079240475052776e-07 & 9.61543204537063425390253166033986e-12\\
85   & 7.47460944734835494603150605681252e-07 & 6.65765514720154023428202200838344e-12\\
86   & 6.25076012803183072949714996778214e-07 & 4.51781968589147435681661138798625e-12\\
87   & 5.27545188407840404034972402364704e-07 & 2.97790805796735353056158063859767e-12\\
88   & 4.51005279547543069512840045259275e-07 & 1.87970510537824614741588711602232e-12\\
89   & 3.92371966727686847652422460137122e-07 & 1.10896242549626643385666064787054e-12\\
90   & 3.49224206935226307895172607412510e-07 & 5.83901681870167816802893811429244e-13\\
91   & 3.19719913320901893897552159916508e-07 & 2.47055052842115889755229649961340e-13\\
92   & 3.02547325679113057732806265306883e-07 & 5.99062680825025191284334866467265e-14\\
\end{tabular}
\caption{Data for Pad\'e approximation of the wave with velocity $c = 1.0922851405$, the steepness $H/\lambda=0.14105777885488320816492860225696$, and $y_0 = -0.289784811618456872977429611644$. Parameters of simulations and Pad\'e approximation are $M = 134217728$, $R(y)\simeq 6.14\times 10^{-27}$, $ err_{pade} \simeq 5.43\times 10^{-27}$, and the smallest Fourier harmonic had value $|\hat z_{k_{max}}| \simeq 3.00\times 10^{-31}$.}
\label{tab:c1.0922851405}
\end{center}
\end{table}

\bibliographystyle{jfm}

\bibliography{surfacewaves,lushnikov}

\end{document}